\documentclass[useAMS,usenatbib]{mn2e}

\usepackage{graphicx}
\title[Wide double stars and multiple systems]
{Double stars with wide separations in the AGK3 - II. The 
wide binaries and the multiple systems\footnotemark[1]\thanks{based on observations performed
at the Observatoire de Haute--Provence (CNRS), France}}
\author[J.-L. Halbwachs, M. Mayor and S. Udry]{J.-L. Halbwachs$^{1}$\thanks{E-mail:
jean-louis.halbwachs@astro.unistra.fr}, M. Mayor$^{2}$ and S. 
Udry$^{2}$\\
$^{1}$Observatoire astronomique de Strasbourg, Universit\'e de Strasbourg, CNRS, UMR 7550,
11 rue de l'Universit\'{e}, 67000 Strasbourg, France\\
$^{2}$Observatoire Astronomique de l'Universit\'e de Gen\`eve, 51, chemin des maillettes,
CH--1290 Sauverny, Switzerland}
\begin{document}

\date{Accepted 2016 October 14. Received 2016 October 13; in original form 2016 August 29}

\pagerange{\pageref{firstpage}--\pageref{lastpage}} \pubyear{2016}

\maketitle

\label{firstpage}

\begin{abstract}
A large observation programme was carried out to measure the radial velocities of the components of
a selection of common proper motion (CPM) stars, in order to select the physical binaries. Eighty
wide binaries (WBs) were detected, and 39 optical pairs were identified. 
Adding CPM stars with separations close
enough to be almost certain they are physical, a ``bias-controlled'' sample of 116 wide binaries
was obtained, and used to derive the distribution of separations from 100 to 30,000 au. The distribution
obtained doesn't match the log-constant distribution, but is in agreement with the log-normal
distribution.
The spectroscopic binaries detected among the WB components were used to derive statistical
informations about the multiple systems.
The close binaries in WBs seem to be similar to those detected in other field stars.
%{\bf 
As for the WBs, they seem to obey the log-normal distribution of periods.
The number of quadruple systems is in agreement with the ``no correlation'' hypothesis;
this indicates that an environment conducive to the formation of WBs doesn't favor the
formation of subsystems with periods shorter than 10 years.
%}

\end{abstract}

%===========================================================================================  

\section{Introduction}

%In a previous paper \citep[Paper I hereafter]{CPM1}, we have presented a programme dedicated
%to the selection of a sample of binaries with large separations. 

Binary stars cover a very wide range of separations, from a few stellar radii to thousands
of astronomical units. 
The binaries with orbital period comparable to a human lifetime are
detected from variations of quantities such as the radial velocity or the position
of the components, but the wide binaries (WBs) are more hazardous to identify. Both components look
like single stars with constant velocities, and statistical criteria are used to infer whether they 
can be bounded by gravitation.

The origin of WBs is still debated, since they can't be obtained from the simple fragmentation of
a collapsing cloud of gas and dust. Other processes were proposed, as gravitational perturbations
from passing stars and giant molecular clouds \citep{WeinbergShaWass,Jiang10}, star cluster dissolution
\citep{Kouwenhoven} or dynamical evolution of triple systems \citep{ReipurthMikko12}. 
This motivates investigations about the statistical properties of these systems, which are
assumed to contain the signature of their formation \citep[see the review by][and references therein]{DucheneKraus13}.

The components of WBs are not all single stars, but some of them are binaries with closer separation. The properties of
the multiple systems, and the possible correlation between the parameters of the WBs and those of the
inner binaries, are also clues to understand the origin of these objects. In addition to researches mentioned in
\citet{DucheneKraus13}, these questions were investigated recently by \citet{Toko14,Toko15}, and by \citet{Riddle15}.
They estimated that the frequency of inner binaries is the same among the primary and among the secondary components
of the WBs, but that the frequency of quadruple systems is too large to assume that the presence of inner binaries is not
correlated. If this is confirmed, the formation of the inner binaries could results from the environment where the WB was formed.

In this paper, we provide a new sample of WBs to derive their statistical properties.
Since the number of optical stars is growing as the square of the apparent separations,
the selection of physical WBs is based on more parameters. One of us \citep[H86 hereafter]{Halbwachs86} used
the proper motion to select 439 common proper motion
(CPM) pairs from the reduction of the second and of the third Astronomischen Gesellschaft Katalog (AGK2/3) by
\citet{Lacroute74}. These wide pairs were listed in two tables, according to the time $T=\theta/\mu$,
where $\theta$ is the apparent separation and $\mu$ the proper motion. Therefore, $T$ is the time
it takes for the proper motion to 
displace the components over a distance equal to their separation. The first table contained 326 pairs of stars with
$T < 1000$~years, and the second one contained 113 pairs with $T$ between 1000 and 3500 years.
To avoid any confusion between the tables in H86 and those in the present article, the tables I and II of
H86 are called, respectively, ``list~1'' and ``list~2'' hereafter.
The rate of optical pairs was estimated around 1~\% in list~1, and 40~\% in list~2. 

An observational programme was initiated in 1986 in order to measure the radial velocities (RVs) of the stars,
and to select the physical binaries. However, several WBs are in fact multiple systems including
close binaries (CBs), and the RVs of 66 stars from the two lists appeared to be variable, or possibly variable. These 66
stars were observed over about 20 years, leading to the derivation of the orbital elements of 52 short-period
spectroscopic binaries (SBs), and the confirmation of periods of a few decades for 11 other stars. The status of the
3 remaining stars is uncertain, and the variability of their RV is questionable. These results
were published in \citet[Paper~I hereafter]{CPM1}.

The present paper is the continuation of Paper I. It is organized as follows: The observation programme
and the mean RVs of the component stars are presented in Section~\ref{sec:RV}. The selection of the
physical binaries is in Section~\ref{sec:CORAVELWB}. 
A bias-controlled sample of WBs with unevolved components is presented in Section~\ref{sec:BCMSWB}.
This sample is used to derived the frequency of WBs in
Section~\ref{sec:stat}. Some properties of the multiple systems are investigated in Section~\ref{sec:multi}. 
Section~\ref{sec:conclusion} is the conclusion.
%\citealt{Tohline02}).

%\citet{Toko06},

%===============================================================================================

\section{The radial velocity measurements}
\label{sec:RV}

The observational programme was detailed in Paper I. In summary, we used the spectrovelocimeter
CORAVEL \citep[for COrrelation RAdial VELocities,][]{Baranne79}, that was installed on the Swiss 1-m telescope at the Observatory
of Haute-Provence (OHP) until its decommissioning in 2000. Our list of targets contained 266 stars from 
the two lists of H86. Since the probability to be physical is much larger for a pair in list~1 than in list~2,
only 90 targets were chosen among the 326 pairs of list~1: the systems with $T$ shorter than 350 years were
discarded, since the expected number of optical systems among them is only 0.5, leading to a frequency of 0.2~\%;
we discard also the pairs with both components fainter than the completeness limit of the AGK2/3, which is $m_{pg}=9.1$~mag.
In the same time, the 113 pairs in list~2 provided 176 targets.
Due to the limitations of CORAVEL, 
these stars are generally as late as or later than F5, but a few Am-type stars were also observed.
The precision of CORAVEL was around 0.3 km~s$^{-1}$ for slow rotators with spectral types around K0, but
around 1 km~s$^{-1}$ for Am stars.

The detection of the SBs was based on the $P(\chi^2)$ test at the 1 percent threshold. The measurements of
the 66 stars suspected to be variable were presented in Paper I. The 200 remaining received a total of 
1473 RV measurements, but the number of measurements is not the same for every stars: 3 early-type stars
were observed only once, and another one received only 2 measurements. At the opposite, the RV of 52 stars
was obtained at least 10 times, usually because the other component of the pair had a variable RV.
Among the 200 stars not discussed in Paper I, 98 were observed at least 5 times. The median time-span
of the observations is 1720 days.

\begin{table*}
 \centering
% \begin{minipage}{125mm}
  \caption{Sample of the RV catalogue. The headers of the first stars, beginning with the number of the list in H86,
the sequentiel number of the pair, and a capital letter designating the component; in the catalogue file, each
header is immediately followed by $N_{meas}$ measurements. The full catalogue is available as Supporting Information with
the electronic version of the article.}
  \begin{tabular}{@{}lclccccccc@{}}
  \hline
CPM          &   AG     & HIP/HD/BD/AG & $B-V$& $N_{meas}$ & $\bar{V}$ & $\sigma_{\bar{V}}$ & $P(\chi^2)$ & \multicolumn{2}{c}{J2000 coordinates} \\
             &          &              & (mag)&            &(km s$^{-1}$)& (km s$^{-1}$)    &             & (HH mm s.sss) &  ($\pm$DD mm s.ss)  \\
  \hline
2 $\;\;$  1A & +17    8 & HIP 493 & 0.60 &  7 & -45.277 & 0.139 & 0.679  &    00 05 54.748 & +18 14 06.02\\
2 $\;\;$  1B & +17    9 & HIP 495 & 0.92 &  4 & -44.973 & 0.189 & 0.956  &    00 05 55.473 & +18 04 33.14\\
2 $\;\;$  3A & +24   37 & BD+23 45& 1.40 &  4 & -53.179 & 0.161 & 0.744  &    00 22 42.473 & +24 34 08.34\\
2 $\;\;$  3B & +24   36 & BD+23 44& 0.58 &  4 & -20.225 & 0.244 & 0.670  &    00 22 37.298 & +24 33 45.75\\      
  \hline
\label{tab:header}
\end{tabular}
%\end{minipage}
\end{table*}

The RV catalogue is in electronic form, in one plain text file. The format is the same as for the RV
catalogue in Paper I.
For each star, the RV measurements are preceded by a header, providing the number of RV measurements and the mean
RV. The format of the headers is presented in Table~\ref{tab:header}, and that of the measurements
is in Table~\ref{tab:mesRV}.

\begin{table}
 \centering
% \begin{minipage}{140mm}
  \caption{Sample of the RV catalogue. The measurements of star 2:1A, following the header.}
  \begin{tabular}{@{}lrlc@{}}
\hline
JD   & \multicolumn{1}{c}{$V_R$}  &  $\sigma_{RV}$ & T \\
+2\,400\,000 & (km~s$^{-1}$) & (km~s$^{-1}$) & \\
\hline
46429.272  & -45.28 & 0.41  &    C \\
46729.412  & -45.05 & 0.39  &    C \\
46770.322  & -44.99 & 0.38  &    C \\
46814.230  & -45.07 & 0.37  &    C \\
47364.626  & -45.06 & 0.39  &    C \\
48138.550  & -45.54 & 0.37  &    C \\
48881.719  & -45.70 & 0.31  &    C \\
\hline
\label{tab:mesRV}
\end{tabular}
%\end{minipage}
\end{table}

Taking into account the SBs in Paper I, the mean RVs are available for both components of 119 CPM systems. 
The selection of the physical pairs is discussed in the next section.

%===============================================================================================

\section{Selection of the wide binaries observed with CORAVEL}
\label{sec:CORAVELWB}

We consider now the 119 CPM systems observed with CORAVEL.
The selection of the binary stars is essentially based on the difference between the RVs of the components. 
In addition, the trigonometric parallax of 131 stars was measured thanks to the Hipparcos satellite
\citep[for HIgh Precision Parallax COllection Satellite;][]{ESA97,vanLeeuwen2007}, including both
components of 44 pairs. Therefore, the binaries are selected on the basis of the RVs for the 119
pairs, but also on the basis of the parallaxes for 44 of them. This last criterion leads to the
detection of only 2 optical pairs with compatible RVs: 2:87 and 2:99.

The compatibility of the RVs is verified as follows:  
When both components have the same RV in reality, 99~\% of the pairs are expected to satisfy the condition
$\| \Delta V \| < 2.575 \; \sigma_{\Delta V}$, where $\Delta V$ is the RV difference and $\sigma_{\Delta V}$ its
uncertainty. However, two additional tolerances are added, as explained hereafter: 

\begin{itemize}
\item
We take into account the contribution of the orbital motion. This term cannot be derived exactly, since
we don't know the orbital parameters, but it is always less than the escape velocity corresponding to
a parabolic motion. We add then:

\begin{equation}
V_{escape} = 42.23 \times \sqrt{\frac{{\cal M}_A + {\cal M}_B}{s}}
\label{eq:Vesca}
\end{equation}

\noindent
where ${\cal M}_A$ and ${\cal M}_B$ are the masses of the components, in solar units, and $s$ the projection of the separation
between the components, in au. The derivation of ${\cal M}_A$, ${\cal M}_B$, and $s$ is explained later, in the present section.
The coefficient 42.23 is used to derive $V_{escape}$ in km~s$^{-1}$. For $s$ 
larger than 1000~au, $V_{escape}$ is less than about 2~km~s$^{-1}$.

\item
We add an allowance to take into account the difference between the spectral types of the components. Thanks to the correction
of the RVs related to the $(B-V)$ color indices, this error is rather small, and an additional tolerance
of 0.11~km~s$^{-1}$ is assumed. This term lead to the selection of 5 additional binaries.

\end{itemize}

In addition to the apparent separation between the components, three parameters are needed to derive $V_{escape}$ according to
equation~\ref{eq:Vesca}: the masses of the components, and the distance of the system. 
To derive them, the spectral types and the $B$ and $V$ magnitudes are obtained from the Simbad database. When $B$ is missing
in Simbad, it is derived from the Tycho $B_T$ and $V_T$ magnitudes \citep{ESA97}.
The mass of each component is derived from the spectral type, but the
luminosity class is often missing. When the trigonometric parallax is known, the luminosity class is
derived from the absolute magnitude. Otherwise, we assume that both components are dwarfs when the spectral type of the 
brightest one is earlier than that of the other, or when the
$(B-V)$ color index of the brightest one is the smallest. We don't take into account the luminosity class Iab-b
of the secondary component of system 1:31, since the trigonometric parallax of this star ($13.85 \pm 2.46$~mas) 
is in good agreement with a main-sequence star,
but certainly not with a supergiant star.
When both components are nearly identical,
we assume that they are dwarfs when the giant luminosity class would lead to a tangential velocity larger than
100~km s$^{-1}$. Thanks to these assumptions, it was possible to assign a mass, but also a distance, to all stars.

We select finally 80 WBs, which are listed in Table~\ref{tab:WBCora}. In this table, the
component ``A'' is the same as in H86; therefore, it is usually the brightest one, but with
a few exceptions.
$\varpi$ is the trigonometric parallax of the system when it was measured for at least one component;
when it was measured for both, the mean value is given. The distance $D$ was derived from the trigonometric
parallax when its accuracy is better than 20~\%; otherwise, it was derived from the spectroscopic parallax.
The separation between the components was converted in the projected separation, $s$, in au.
The flag in the last column is assigned in ascending order: it is
``1'' when the system is member of an open cluster, otherwise it is ``2'' when the components of the system are not 
main-sequence (MS) stars later than F5, 
otherwise, it is ``3'' when the components are both fainter than $m_{pg} = 9.1$~mag; lastly it is ``4''
for one system (2:49) which is in fact a faint extension of another one (2:48). 
Therefore, a blank flag indicates a field WB system of late-type MS stars with at least one
component brighter than the
completeness limit of the AGK2/3.
The allotment process of these flags is described in
Section~\ref{sec:BCMSWBCoravel} hereafter.

\begin{table*} 
\caption{\label{tab:WBCora} The 80 WBs detected with CORAVEL radial velocities. The flag "f" indicates the
systems rejected from the bias-controlled sample of MS binaries; it is ``1'' when the system belongs to an open cluster,
``2'' when the primary component is earlier than F5 or far from the main sequence, ``3'' when the components are both fainter than $m_{pg}$=9.1~mag
and ``4'' when the system contains the faintest component of a triple CPM system.}
\scriptsize
\begin{tabular}{l|lrrl|lrrl|rrrrc}
\hline
%    & \multicolumn{4}{c}{``A'' component} & \multicolumn{4}{c}{``B'' component} & &&&& \\ 
%CPM & AG & $V$  &  $B-V$ & SpT &  AG & $V$ &  $B-V$ & SpT & $RV_A-RV_B$ & $\varpi$  & $D$ & $s$ & f \\
 CPM & AG(A) & $V_A$  &  $B_A-V_A$ & SpT(A) &  AG(B) & $V_B$ &  $B_B-V_B$ & SpT(B) & $RV_A-RV_B$ & \multicolumn{1}{c}{$\varpi$}  & $D$ & $s$ & f \\
    &       & (mag)   &  (mag)      &         &         & (mag)  &  (mag)      &         &(km s$^{-1}$) & \multicolumn{1}{c}{(mas)} & (pc)& (au)&   \\
\hline
2:  1 & +17    8 &  7.45 &  0.56 &F8        & +17    9 &  8.58 &  0.93 & K0       &$ -0.30\pm  0.23$ &$ 26.93  \pm   0.51 $ &   37 & 21296 &   \\
2:  4 & +33   51 &  5.88 &  1.13 &K1III     & +33   52 &  9.28 &  0.61 & F8       &$ -0.71\pm  0.36$ &$  8.61  \pm   0.36 $ &  116 &  6536 & 2 \\
2:  7 & +31   55 &  8.78 &  0.88 &G5        & +31   54 &  9.44 &  1.04 & G5       &$ -0.01\pm  0.18$ &                      &   61 &  3343 & 3 \\
1:  9 & +12   88 &  8.91 &  0.54 &F8        & +12   89 &  9.18 &  0.62 & G0       &$ -0.40\pm  0.33$ &$  7.66  \pm   0.82 $ &  131 &  5789 &   \\
2:  8 & -01   87 &  9.16 &  0.53 &F7V       & -01   90 & 10.11 &  0.92 & G5       &$ -2.10\pm  0.62$ &                      &   98 &  9643 & 3 \\
1: 17 & +11  131 &  8.32 &  0.58 &F8        & +11  132 & 10.22 &  0.68 & K0       &$ -0.93\pm  0.33$ &$ 10.56  \pm   0.66 $ &   95 &  5148 &   \\
1: 18 & +31  132 &  7.98 &  0.73 &G0        & +31  130 &  8.02 &  0.61 & F8       &$ -0.54\pm  0.13$ &$ 16.19  \pm   0.58 $ &   62 &  3532 &   \\
2:  9 & +37  140 &  7.52 &  0.53 &F8        & +37  138 &  8.95 &  0.76 & K0       &$ -0.35\pm  0.25$ &$ 18.71  \pm   0.64 $ &   53 & 13747 &   \\
1: 20 & +60  173 &  8.55 &  0.52 &F6V       & +60  175 &  9.07 &  0.52 & F6V      &$ -0.11\pm  0.33$ &                      &  100 &  4437 &   \\
1: 27 & +30  199 &  7.74 &  0.47 &F5        & +30  198 &  9.92 &  0.71 & G0       &$ -0.79\pm  0.38$ &$ 11.62  \pm   1.01 $ &   86 &  5668 &   \\
1: 31 & +58  244 &  8.07 &  0.42 &F5        & +58  243 &  9.94 &  0.71 & M0Iab-b$^a$  &$ -1.25\pm  0.31$ &$ 11.17  \pm   1.11 $ &   90 &  5103 &   \\
2: 21 & +20  280 &  6.40 &  1.26 &K2        & +20  279 &  8.72 &  0.46 & F5       &$  1.75\pm  2.72$ &$  6.64  \pm   0.51 $ &  151 & 18365 & 2 \\
2: 22 & +38  399 &  7.92 &  0.39 &F0        & +38  398 &  9.72 &  0.66 & G0       &$ -0.09\pm  0.30$ &$ 12.20  \pm   0.75 $ &   82 & 11189 & 2 \\
2: 33 & +63  319 &  8.30 &  0.79 &G7V       & +63  318 &  9.86 &  1.04 & K2V      &$ -0.96\pm  0.23$ &$ 15.93  \pm   0.67 $ &   63 & 13167 &   \\
2: 36 & +16  384 &  4.77 &  0.18 &A6IV      & +16  385 &  6.51 &  0.41 & F2       &$  6.61\pm  4.91$ &$ 23.50  \pm   0.25 $ &   43 & 10637 & 1 \\
2: 38 & +18  367 &  7.12 &  0.67 &G5        & +18  369 &  9.86 &  1.18 & K7       &$ -0.38\pm  0.20$ &$ 22.81  \pm   0.82 $ &   44 &  6227 &   \\
2: 40 & +29  536 &  6.66 &  0.52 &F8V       & +29  534 &  8.40 &  0.74 & G0       &$ -0.19\pm  0.24$ &$ 18.84  \pm   0.58 $ &   53 &  3649 &   \\
2: 41 & +27  482 &  6.96 &  0.48 &F8V       & +27  481 &  9.25 &  0.94 & G5       &$ -0.64\pm  0.14$ &$ 25.05  \pm   0.67 $ &   40 & 12572 &   \\
2: 42 & +26  477 &  9.26 &  1.57 &K5        & +26  478 &  9.11 &  1.39 & G5       &$ -0.40\pm  0.27$ &$  4.33  \pm   1.14 $ &   46 &  3643 & 2 \\
2: 44 & +44  623 &  6.73 &  0.56 &F8        & +44  621 &  9.10 &  0.94 & ~        &$ -0.11\pm  0.28$ &$ 30.20  \pm   0.44 $ &   33 &  6361 &   \\
1: 82 & +15  742 &  7.72 &  0.64 &G0V       & +15  743 &  7.72 &  0.53 & F8       &$ -0.23\pm  0.36$ &$ 21.62  \pm   0.58 $ &   46 &  7958 &   \\
2: 48 & +13  732 &  7.31 &  0.45 &F5        & +13  734 &  7.98 &  0.54 & F8       &$  1.40\pm  0.15$ &$ 21.64  \pm   0.65 $ &   46 &  5149 &   \\
2: 49 & +13  734 &  7.31 &  0.45 &F5        & +13  733 &  9.21 &  0.94 & G5       &$  0.56\pm  0.19$ &$ 22.02  \pm   0.96 $ &   45 &  5075 & 4 \\
1: 90 & +47  746 &  8.48 &  0.50 &F7V       & +47  747 &  9.77 &  0.70 & G7V      &$ -0.59\pm  0.96$ &$ 12.03  \pm   1.69 $ &   83 &  3175 &   \\
1: 91 & +57  696 &  7.56 &  0.42 &F2        & +57  695 &  9.27 &  0.58 & ~        &$ -0.50\pm  0.26$ &$ 14.24  \pm   0.61 $ &   70 &  4083 & 2 \\
1: 93 & +32  859 &  8.35 &  0.66 &G5        & +32  858 &  9.25 &  0.71 & K2       &$ -1.33\pm  0.15$ &$ 17.30  \pm   1.77 $ &   58 &  6206 &   \\
1:106 & +27  979 &  8.27 &  0.53 &G0V       & +27  978 &  8.26 &  0.50 & G0V      &$  0.15\pm  0.31$ &$ 22.51  \pm   0.85 $ &   44 &  2273 &   \\
2: 55 & +51  744 &  6.11 &  0.42 &F3V       & +51  745 &  7.80 &  0.74 & G5       &$ -1.24\pm  0.16$ &$ 36.08  \pm   0.41 $ &   28 &  6399 & 2 \\
1:112 & -02  517 &  4.60 &  0.46 &F5V       & -02  518 &  7.15 &  0.87 & K0       &$ -1.16\pm  0.20$ &$ 57.69  \pm   2.14 $ &   17 &  1137 &   \\
1:114 & +14 1033 &  9.01 &  0.72 &G5        & +14 1032 &  8.74 &  0.64 & G0       &$ -0.19\pm  0.14$ &                      &   62 &  5094 &   \\
2: 56 & -00 1459 &  6.71 &  1.06 &K0III     & -00 1460 & 10.50 &  0.67 & K        &$ -0.34\pm  0.31$ &$  9.06  \pm   0.52 $ &  110 & 12061 & 2 \\
2: 57 & +33 1010 &  7.98 &  0.47 &F8        & +33 1011 &  8.72 &  0.54 & G        &$ -0.70\pm  0.32$ &$ 11.35  \pm   0.70 $ &   88 & 18122 &   \\
2: 58 & +21 1119 &  8.80 &  0.38 &F8        & +21 1120 &  9.15 &  0.40 & F8       &$  0.37\pm  0.39$ &$  4.80  \pm   0.85 $ &  208 & 40785 &   \\
1:127 & +56  797 &  4.83 &  0.49 &F8V       & +56  796 &  8.76 &  1.34 & M0       &$ -0.26\pm  0.13$ &$ 78.26  \pm   0.28 $ &   13 &  1569 &   \\
1:130 & +46  852 &  5.20 &  0.31 &F5III     & +46  854 &  7.28 &  0.58 & F9V      &$  0.87\pm  5.34$ &$ 26.93  \pm   0.23 $ &   37 & 10671 &   \\
1:140 & +43 1013 &  7.18 &  0.53 &F8        & +43 1012 &  8.24 &  0.60 & G5       &$ -1.57\pm  0.27$ &$ 21.12  \pm   0.47 $ &   47 &  6384 &   \\
1:141 & +02 1472 &  8.32 &  1.02 &K0/1III   & +02 1473 & 10.21 &  0.50 & F7V      &$ -0.55\pm  0.16$ &$  4.55  \pm   0.90 $ &  220 & 14048 & 2 \\
2: 65 & +41 1063 &  8.05 &  0.47 &F8        & +41 1064 &  9.10 &  0.65 & G        &$ -0.02\pm  0.25$ &$ 12.54  \pm   1.15 $ &   80 & 10671 &   \\
1:156 & +20 1296 &  4.54 &  0.52 &A7V+...   & +20 1295 &  8.96 &  0.59 & G5       &$ -2.11\pm  1.02$ &$ 14.02  \pm   0.23 $ &   71 &  5315 & 1 \\
1:158 & +09 1459 &  7.34 &  0.96 &K0        & +09 1458 &  7.87 &  0.47 & K0III-IV &$ -0.26\pm  0.18$ &$ 10.25  \pm   0.53 $ &   98 &  3031 & 2 \\
2: 66 & +63  648 &  9.21 &  0.59 &G         & +63  650 & 10.03 &  0.76 & G5       &$  1.12\pm  0.31$ &                      &   92 & 13421 & 3 \\
2: 70 & +36 1135 &  8.10 &  0.46 &F8V       & +36 1139 &  8.20 &  0.38 & F3       &$ -0.12\pm  0.19$ &$  9.33  \pm   0.93 $ &  107 & 25734 &   \\
2: 71 & +73  326 &  9.79 &  0.71 &G0        & +73  327 & 10.03 &  0.65 & F8       &$ -0.23\pm  0.39$ &                      &  116 &  9018 & 3 \\
2: 73 & +20 1355 &  8.43 &  0.79 &K0V       & +20 1356 &  8.80 &  0.95 & K3V      &$  0.05\pm  0.13$ &                      &   30 & 12354 & 1 \\
2: 74 & +25 1366 &  9.10 &  0.87 &K0IV      & +25 1369 &  9.92 &  1.20 & K5IV-V   &$ -0.06\pm  0.16$ &$ 26.63  \pm   0.97 $ &   38 & 13993 & 1 \\
1:175 & +21 1322 &  4.91 &  0.89 &G5III     & +21 1323 &  9.36 &  0.83 & ~        &$ -1.38\pm  0.16$ &$ 11.52  \pm   0.87 $ &   87 &  2456 & 1 \\
2: 76 & +67  574 &  6.52 &  1.15 &K2III     & +67  570 &  6.96 &  1.15 & K2III    &$ -0.41\pm  0.18$ &$  8.58  \pm   0.27 $ &  117 & 20855 & 2 \\
1:190 & +10 1635 &  6.49 &  1.04 &K0III     & +10 1636 &  8.97 &  0.44 & F6V      &$ -0.50\pm  0.24$ &$  8.38  \pm   0.49 $ &  119 &  8364 & 2 \\
1:194 & +38 1289 &  5.51 &  1.04 &K0III     & +38 1288 &  8.86 &  0.53 & F8V      &$ -0.67\pm  0.23$ &$ 10.29  \pm   0.33 $ &   97 &  6950 & 2 \\
1:196 & +26 1388 &  6.95 &  0.44 &F6V       & +26 1389 &  8.83 &  0.77 & K0       &$ -0.12\pm  0.27$ &$ 22.72  \pm   0.50 $ &   44 &  4269 &   \\
2: 88 & +30 1364 &  7.69 &  0.62 &G0        & +30 1366 &  7.95 &  0.62 & G2IV     &$ -0.19\pm  0.19$ &$ 18.93  \pm   0.50 $ &   53 & 16088 &   \\
2: 89 & +83  385 &  9.55 &  0.60 &K0        & +83  383 & 10.71 &  0.70 & F8       &$ -3.54\pm  2.66$ &$  1.00  \pm  11.00 $ &  136 &  8117 & 2 \\
1:216 & +33 1351 &  3.49 &  0.95 &G8IV      & +33 1352 &  7.81 &  0.58 & G0Vv     &$ -0.59\pm  0.20$ &$ 26.79  \pm   0.16 $ &   37 &  3913 & 2 \\
1:237 & +08 1965 &  6.97 &  0.88 &G5        & +08 1966 &  8.26 &  0.55 & G5       &$  0.18\pm  0.19$ &$ 13.32  \pm   0.54 $ &   75 &  4414 & 2 \\
2: 91 & +69  664 &  8.01 &  0.26 &F0        & +69  663 &  8.00 &  0.40 & F5       &$  3.17\pm  5.56$ &$  9.03  \pm   0.48 $ &  111 & 16267 & 2 \\
1:241 & +06 1960 &  6.58 &  0.88 &G5        & +06 1958 & 10.30 &  1.04 & K3       &$  0.25\pm  0.32$ &$ 21.94  \pm   0.57 $ &   46 &  7427 &   \\
2: 92 & +05 2201 &  9.73 &  0.84 &G5        & +05 2202 & 10.90 &  0.52 & F8       &$  0.18\pm  0.29$ &                      &   98 &  5800 & 3 \\
1:242 & +47 1214 &  7.88 &  1.04 &K3V       & +47 1216 &  7.83 &  0.98 & K3V      &$ -0.50\pm  0.12$ &$ 55.54  \pm   0.39 $ &   18 &  2033 &   \\
1:246 & +86  233 &  8.35 &  0.39 &F2        & +86  232 &  9.20 &  0.50 & G0       &$  0.01\pm  0.22$ &$  8.73  \pm   0.49 $ &  115 &  3578 & 2 \\
2: 94 & +46 1275 &  8.27 &  0.50 &F5        & +46 1274 & 10.28 &  0.78 & G5       &$ -0.22\pm  0.18$ &                      &  100 & 19920 &   \\
2: 95 & +63  904 &  7.67 &  0.52 &F9V       & +63  903 &  8.39 &  0.73 & G0       &$ -0.17\pm  0.21$ &$ 22.12  \pm   0.36 $ &   45 &  8750 &   \\
1:258 & +28 1759 &  8.10 &  0.52 &F8        & +28 1761 &  8.22 &  0.53 & F8       &$  0.19\pm  0.13$ &$ 12.46  \pm   0.61 $ &   80 &  4472 &   \\
1:261 & +45 1487 &  8.33 &  0.48 &F8IV      & +45 1486 &  9.29 &  0.51 & G0.5IV   &$ -0.75\pm  0.29$ &$  9.48  \pm   0.67 $ &  105 &  3100 &   \\
2:100 & +03 2375 &  8.30 &  0.42 &F2/3V     & +03 2377 &  9.16 &  0.49 & G0       &$ -0.61\pm  0.31$ &                      &   87 &  7649 & 2 \\
1:264 & +34 1826 &  7.25 &  0.45 &F5        & +34 1825 & 10.16 &  0.94 & K0       &$ -3.46\pm  0.81$ &$ 15.02  \pm   0.50 $ &   67 &  2191 &   \\
1:280 & -02 1185 &  6.70 &  0.47 &F7V       & -02 1184 &  7.48 &  0.48 & F6V      &$  0.11\pm  0.13$ &$ 21.58  \pm   0.41 $ &   46 &  2763 &   \\
2:103 & +17 2205 &  9.04 &  0.49 &G0        & +17 2206 &  9.79 &  0.59 & G0       &$  0.34\pm  0.37$ &                      &   92 & 11631 & 3 \\
1:282 & +14 2238 &  8.59 &  0.43 &G0        & +14 2237 & 10.26 &  0.43 & G5       &$ -1.42\pm  0.29$ &$  8.10  \pm   0.96 $ &  123 &  9234 &   \\
1:283 & +80  431 &  5.96 &  0.92 &K0III+..  & +80  432 &  8.66 &  0.59 & G5       &$ -0.45\pm  0.23$ &$ 15.48  \pm   0.21 $ &   65 & 13832 & 2 \\
1:284 & +32 1986 &  8.17 &  0.53 &F6V       & +32 1987 &  8.69 &  0.62 & F9V      &$ -1.04\pm  0.34$ &                      &   73 &  3908 &   \\
2:106 & +41 2028 &  6.60 &  0.91 &G0        & +41 2027 &  8.81 &  0.58 & G0       &$ -0.34\pm  0.31$ &$ 14.83  \pm   0.38 $ &   67 & 31003 &   \\
2:107 & +57 1428 &  8.13 &  0.54 &G5        & +57 1427 &  8.63 &  0.53 & K0       &$  0.22\pm  0.25$ &$ 10.56  \pm   0.69 $ &   95 &  7951 &   \\
1:300 & +18 2249 &  8.48 &  0.44 &F8        & +18 2248 & 10.46 &  0.44 & G5       &$ -0.78\pm  0.23$ &                      &  129 &  8494 &   \\
1:307 & +72  613 &  7.58 &  0.39 &F5        & +72  614 &  8.41 &  0.42 & F5       &$ -1.27\pm  0.44$ &$ 11.81  \pm   0.44 $ &   85 &  3616 &   \\
2:110 & +29 2849 &  8.47 &  0.38 &F5V       & +29 2850 & 10.38 &  0.67 & K0       &$  0.03\pm  0.36$ &$  8.40  \pm   1.00 $ &  119 & 20744 &   \\
\end{tabular}
\flushleft \bf
$^a$ the luminosity class is ignored, since it doesn't match the magnitude and the trigonometric parallax of the star; see explanations in the text.
\end{table*}

\addtocounter{table}{-1}
\begin{table*}
\caption{(Continued)}
\scriptsize
\begin{tabular}{llrrllrrlrrrrc}
\hline
CPM & AG(A) & $V_A$  &  $B_A-V_A$ & SpT(A) &  AG(B) & $V_B$ &  $B_B-V_B$ & SpT(B) & $RV_A-RV_B$ & $\varpi$  & $D$ & $s$ & f \\
    &       & (mag)   &  (mag)      &         &         & (mag)  &  (mag)      &         &(km s$^{-1}$) & (mas)           & (pc)& (au)&   \\
\hline
1:310 & +67 1041 &  6.98 &  0.42 &F2        & +67 1040 &  7.52 &  0.46 & F5       &$ -0.72\pm  0.61$ &$ 16.81  \pm   0.28 $ &   59 &  4129 & 2 \\
2:111 & +45 2142 &  8.49 &  0.64 &G5        & +45 2141 &  7.88 &  0.57 & G0       &$ -0.21\pm  0.22$ &$ 17.63  \pm   0.70 $ &   57 & 10503 &   \\
2:112 & +59 1650 &  9.70 &  0.58 &K0        & +59 1651 &  9.56 &  0.60 & G0       &$ -0.21\pm  0.30$ &                      &   93 & 10202 & 3 \\
2:113 & +40 2513 &  7.71 &  0.58 &G0        & +40 2512 &  8.15 &  0.64 & G0       &$  0.66\pm  0.27$ &$ 23.21  \pm   0.46 $ &   43 &  4948 &   \\
1:326 & +10 3290 &  8.73 &  0.62 &G0        & +10 3289 &  8.50 &  0.55 & F8       &$ -0.65\pm  0.27$ &$ 12.18  \pm   0.65 $ &   82 &  5212 &   \\
\hline
\end{tabular}
\end{table*}

The 39 remaining pairs are considered
as optical, and their CPM identifications are listed in Table~\ref{tab:optical}.

\begin{table}
\centering
\begin{minipage}{1\columnwidth}
\caption{\label{tab:optical} The 39 optical pairs with
CORAVEL radial velocities for both components. Two of them (2:87 and 2:99) are detected by the Hipparcos
trigonometric parallaxes of the components.}
\begin{tabular}{lccccccr}
\hline
CPM  & CPM & CPM & CPM & CPM & CPM   & CPM & CPM\\
\hline
2:  3 & 2:  5 & 2:  6 & 1: 19 & 2: 10 & 2: 13 & 2: 14 & 2: 15 \\
2: 16 & 2: 17 & 2: 18 & 2: 19 & 2: 46 & 2: 53 & 2: 54 & 2: 59 \\
2: 62 & 2: 63 & 2: 64 & 2: 67 & 2: 68 & 2: 72 & 2: 75 & 2: 78 \\
2: 80 & 2: 81 & 2: 82 & 2: 83 & 1:195 & 2: 84 & 2: 85 & 2: 87 \\
2: 90 & 2: 93 & 2: 99 & 2:101 & 2:105 & 2:108 & 2:109 &       \\
\hline
\end{tabular}
\end{minipage}
\end{table}

%----------------------

\begin{figure}
\includegraphics[clip=,width=\columnwidth]{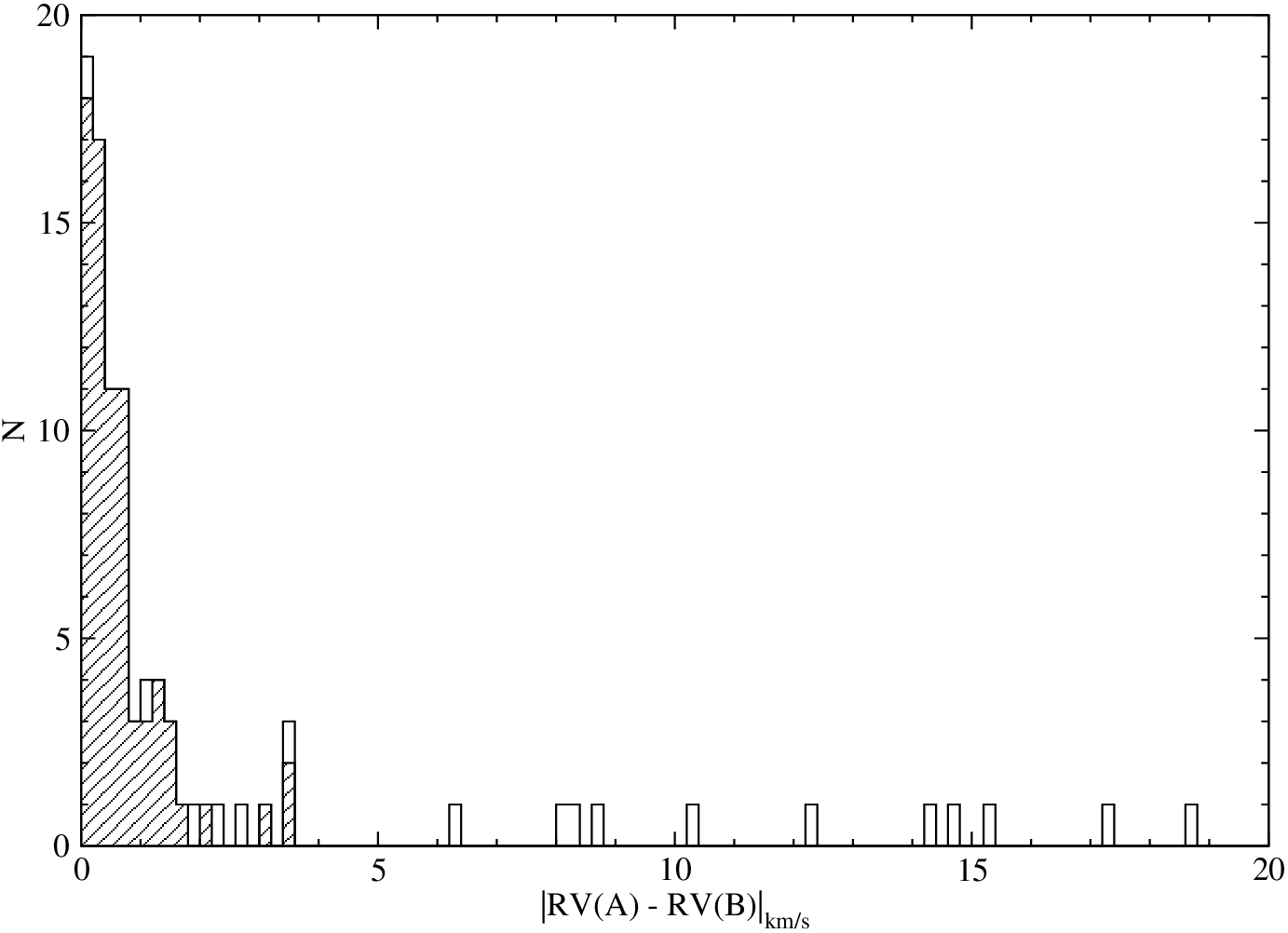}
 \caption{Histogram of the differences of radial velocities for the 119 CPM pairs
with both components observed with CORAVEL. The hatched area refers to the WBs in
Table~\ref{tab:WBCora}.
}
\label{fig:DVR}
\end{figure}

%----------------------------

The distribution of the difference of RVs of the components is in Fig.~\ref{fig:DVR}.
It is worth noticing that we count 8 optical pairs among the stars with a RV difference
between 2 and 10~km~s$^{-1}$, i.e. 1 pair per km~s$^{-1}$. Among the pairs with RV difference
less than 2~km~s$^{-1}$, we count exactly 2 optical pairs, i.e. again 1 pair per km~s$^{-1}$,
although their detection comes from the parallaxes only. 
This indicates that very few optical
pairs, if any, were misclassified as physical binaries. 

%===============================================================================================

\section{The bias-controlled sample of main-sequence wide binaries}
\label{sec:BCMSWB}

The statistical properties of WBs may be derived only when the selection effects are
properly taken into account. This requires a sample which is as complete as possible, and for
which the probability detection may be estimated. The WBs listed in Table~\ref{tab:WBCora}
do not satisfy these two conditions: the selection effects cannot be estimated for all of them,
and they are not a complete sample, since we didn't observe many close pairs of list~1. 
Therefore, only a part of them are selected in Section~\ref{sec:BCMSWBCoravel}
hereafter, and a supplementary list of WBs is still added in Section~\ref{sec:suppList}.

\begin{figure}
\includegraphics[clip=,width=\columnwidth]{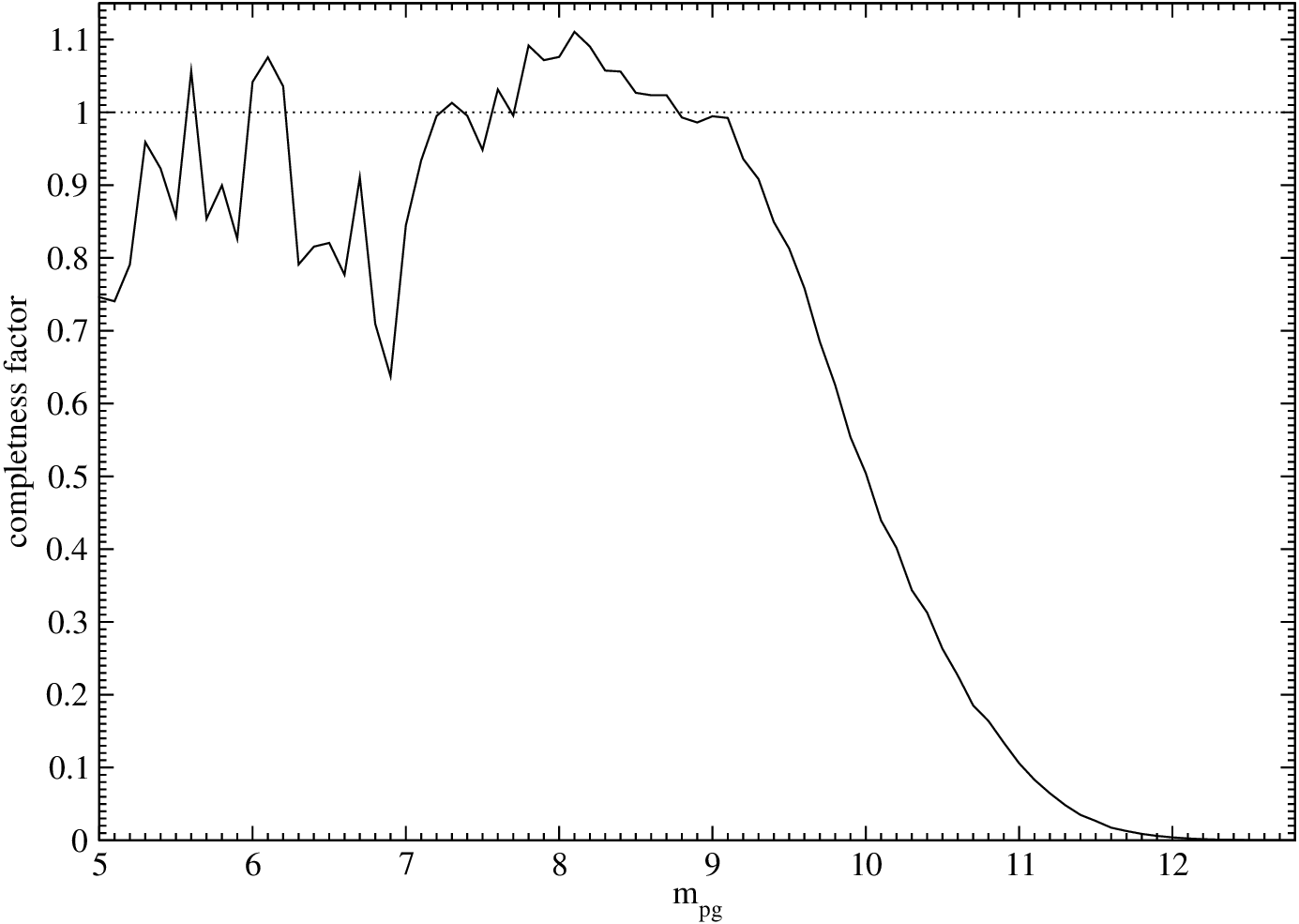}
 \caption{The completeness factor of the AGK2/3 as a function of the photographic magnitude, when
the distribution in equation~\ref{eq:fmpg} is assumed. For $m_{pg} \leq 9.1$~mag, the deviations
from unity are ignored.
}
\label{fig:probaDet}
\end{figure}

%------------------------------------------------------------------------------

\subsection{The bias-controlled sample of CORAVEL main-sequence wide binaries}
\label{sec:BCMSWBCoravel}

First of all, we reject the systems which probably belong to open clusters. For them, the
fact that the components have the same proper motion, the same radial velocity, and the same parallax
doesn't mean that they are very probably orbiting binaries. We search the systems which are as close as 10 degrees
around the centers of the four most important clusters of the northern hemisphere, which are the
Hyades, the Pleiades, Praesepe and Coma. When the mean RV of a system is that of the cluster, {\it plus} or
{\it minus} 10 km s$^{-1}$, we checked if it is a known member. Five systems
are thus selected: 2:36 belongs to the Hyades,
and 1:156, 2:73, 2:74 and 1:175 belong to the Coma cluster. They are indicated by the flag ``1'' in Table~\ref{tab:WBCora}.

The method used to derive the intrinsic distribution of the projected separations, in Section~\ref{sec:stat},
assumes that the probability distribution of the apparent magnitude of the secondary component is known.
%The derivation of the distribution of the separations, in Section~\ref{sec:stat}, is feasible only when we know
%the probability distribution of the apparent magnitude of the secondary component. 
When the primary component
is a late-type giant or a supergiant star, this function is hazardous to derive,
since it depends on the age of the binary system and on the evolution of both components.
Therefore, we decide to consider only systems of components close to the main sequence, 
according to the procedure applied in Section~\ref{sec:CORAVELWB}.
We also reject the binaries with a primary component earlier than F5. These early-type stars are difficult to
observe with CORAVEL, and it was not possible to observe all of them. 
The systems rejected because of spectral type or of luminosity class
are indicated with the flag ``2'' (except when they are already flagged ``1'').

We have a problem with the pairs of components both
fainter than the photographic magnitude $m_{pg} = 9.1$~mag in the AGK2/3 catalogue; below this value,
the density of stars as a function of $m_{pg}$ is well fitted with the following equation:

\begin{equation}
f(m_{pg}) = 3.084 \; {\rm e}^{1.068 \; m_{pg}}
\label{eq:fmpg}
\end{equation}

\noindent
but the completeness factor drops sharply beyond, as shown in Fig.~\ref{fig:probaDet}. The completeness factor is defined
as the ratio between the number of stars counted in a 0.1-mag bin, and the number predicted by equation~\ref{eq:fmpg}.
For magnitudes brighter than 9.1, the completeness factor is 1, with a variance due to the number of stars per bin.
For magnitudes fainter than 9.1, 
the completeness factor is decreasing, as is the
probability that the star is in the AGK2/3. It is tempting to assume that the probability to include
a pair of stars fainter than 9.1~mag is the product of the completeness factors of both components; unfortunately,
this is wrong, since the detection probabilities of two stars recorded on the same photographic plate
are correlated. For that reason, the WBs with both components fainter than 9.1 mag are discarded, and
they receive the flag ``3''. For the remaining WBs, the completeness factor of the component fainter than
9.1~mag is the probability to include the pair in the AGK2/3.

We also discard the pair 2:49 (flag ``4''), since its primary component 
is also that of 2:48; in order to consider only wide binaries, but not wide triple systems, we keep the pair
with the brightest secondary component.

At the end, we keep only 47 binaries with
RVs observed with CORAVEL. These stars are indicated with a blank flag in Table~\ref{tab:WBCora}.

%-----------------------------------------------------------------------------

\subsection{The supplementary list of main-sequence wide binaries}
\label{sec:suppList}

Many pairs in the lists of H86 were not observed with CORAVEL since their probability to be gravitationally bound is
very high. This applies to the pairs of list~1 with $T<350$~years, and a few others for which the RVs were already well known. 
These systems must be taken into account when their primary component is at least as bright as $m_{pg}=9.1$~mag, is at least as late as
F5, and when the components seems to be close to the MS. They are selected as explained hereafter.

First of all, the same selection criterions as for the systems
observed with CORAVEL are applied. The status of the pair 1:213 is puzzling. On one hand, the components have incompatible
parallaxes in Hipparcos~2 \citep{vanLeeuwen2007}. On the other hand, the parallaxes were compatible in the first Hipparcos reduction,
$T$ is only 40~years, the spectral type of the secondary component matches the absolute magnitude
derived from the first Hipparcos reduction much better than that derived from Hipparcos~2, and, above all, the
difference of the RVs of the components, obtained from Simbad, is only $(0.06 \pm 0.11)$~km~s$^{-1}$. Therefore,
we finally decided to consider this pair as a physical binary.

One more selection criterion must still be added: by including the pairs with $T<350$~years, we include the
pairs with separations close to the resolving power of the instrumentation used to prepare the AGK2/3 catalogue. Therefore,
some close pairs may be included in our list by chance, although they are so close that the probability to detect
the secondary component was small. This may induce a selection effect that will be hazardous to correct. By
security, we prefer to discard the pairs too close to be sure that their detection was complete. We assume that the components
of a close pair are certainly both in the AGK2/3 when the separation is larger than the sum of overestimations of the images of
the components derived in the Appendix, Equation~\ref{eq:RMpg}. This results is discarding the
12 pairs listed in Table~\ref{tab:WBClo}. The supplementary list to the controlled-biased sample of MS binaries is thus obtained. 
It is presented in Table~\ref{tab:WBcomp}. As for the CORAVEL pairs, we ignore the luminosity class of the star when it is obviously erroneous.
This concerns the two following stars: (1) 
1:192B = AG +39 1356, since its proper motion leads to a tangential velocity of about 660 km~s$^{-1}$
when the star is assumed to be a giant, and (2) 1:223A = HIP 75676 has the absolute
magnitude $M_V=4.94$~mag according to Hipparcos data. Both are taken as dwarfs instead of giants.

\begin{table}
\centering
\begin{minipage}{1\columnwidth}
\caption{\label{tab:WBClo} The 12 WBs rejected from the MS bias-controlled sample since the components
are so close that similar systems could have been undetected.}
\begin{tabular}{lccccccr}
\hline
CPM  & CPM & CPM & CPM & CPM & CPM   & CPM & CPM\\
\hline
1: 65 & 1: 69 & 1: 84 & 1: 95 & 1:129 & 1:146 & 1:202 & 1:208 \\
1:265 & 1:288 & 1:293 & 1:320 &       &       &       &       \\
\hline
\end{tabular}
\end{minipage}
\end{table}

The systems with blank flag in Table~\ref{tab:WBCora} and all those in \ref{tab:WBcomp} constitute the 
bias-controlled sample of MS WBs, or, for
brevity, the BC sample hereafter.
This sample contains 116 systems. The primary components
have spectral types between F5 and K3. This range is not much larger than that of the sample studied
by \citet{DM91}, which was F7--G9, or by \cite{Raghavan10}, which was F6--K3.
The distribution of $\log s$ of the BC sample is presented in Figure~\ref{fig:flogs}.

\begin{table*} 
\caption{\label{tab:WBcomp} The 69 WBs of the supplementary list of the BC sample.}
\scriptsize
\begin{tabular}{llrrllrrlrrrrr}
\hline
CPM &  AG(A) & $V_A$  &  $B_A-V_A$ & SpT(A) &  AG(B) & $V_B$ &  $B_B-V_B$ & SpT(B) & $\rho$ & $\varpi$ & $\sigma_\varpi$ & $D$ & $s$  \\
    &        & (mag)   &  (mag)    &        &        & (mag) &  (mag)     &        & (arcsec)   & (mas)    & (mas)           & (pc)& (au)   \\
\hline
1: 14 & +51  118 &  8.09 &  0.45 & F8       & +51  117 &  8.53 &  0.49 & F8       &  14.6 &  13.06 &   3.19 &   66 &   970 \\
1: 29 & -00  223 &  6.87 &  0.62 & G2V      & -00  222 & 10.52 &  1.23 & K5       &  79.0 &  27.06 &   0.58 &   37 &  2919 \\
1: 32 & +28  257 &  7.00 &  0.49 & F5V      & +28  256 &  7.74 &  0.59 & G2V      &  14.4 &  30.46 &   2.58 &   33 &   473 \\
1: 34 & +34  226 &  7.75 &  0.58 & G0IV     & +34  227 &  9.40 &  0.60 & F8       &  10.2 &  11.14 &   1.30 &   90 &   917 \\
1: 35 & +15  205 &  8.88 &  0.63 & G5       & +15  204 &  9.38 &  0.69 & G5       &  34.7 &  17.20 &   1.04 &   58 &  2015 \\
1: 38 & +55  259 &  6.95 &  0.48 & F5V      & +55  258 &  9.58 &  0.77 & K2V      &  21.1 &  18.28 &   0.69 &   55 &  1154 \\
1: 39 & +24  227 &  6.48 &  0.38 & F6III    & +24  226 &  7.09 &  0.50 & F4V      &  37.9 &  24.19 &   0.45 &   41 &  1567 \\
1: 43 & +38  324 &  5.36 &  0.37 & F6V      & +38  323 &  9.68 &  0.78 & ~        &  14.2 &  14.15 &   0.72 &   71 &  1004 \\
1: 47 & +07  353 &  7.37 &  0.59 & F8       & +07  354 &  7.75 &  0.63 & G0       & 156.0 &  22.50 &   0.73 &   44 &  6933 \\
1: 51 & +63  274 &  6.80 &  0.46 & F5       & +63  275 &  8.20 &  0.83 & G5       &  45.8 &  23.73 &   0.47 &   42 &  1931 \\
1: 53 & +41  396 &  8.78 &  0.15 & K1V      & +41  397 &  8.78 &  0.88 & K2V      &   7.4 &  45.65 &   2.63 &   22 &   162 \\
1: 64 & +07  535 &  8.38 &  0.89 & K0       & +07  536 &  8.20 &  0.83 & K0       &  16.4 &  33.00 &   4.00 &   30 &   496 \\
1: 66 & +63  355 &  7.60 &  0.67 & G0       & +63  354 &  7.90 &  0.74 & G0       &  32.5 &  31.03 &   0.61 &   32 &  1048 \\
2: 43 & +17  485 &  5.00 &  0.52 & F8V      & +17  482 &  7.92 &  1.13 & K4V      & 707.2 &  69.62 &   0.37 &   14 & 10158 \\
1: 70 & +54  477 &  7.53 &  0.64 & G1V      & +54  476 &  9.71 &  1.00 & K3:      &   8.4 &  25.44 &   1.04 &   39 &   329 \\
1: 72 & +53  486 &  6.23 &  0.84 & K1V      & +53  487 &  9.87 &  1.44 & M0.5     &  98.0 &  81.35 &   0.51 &   12 &  1204 \\
1: 75 & -00  757 &  8.74 &  0.61 & G5       & -00  758 & 10.07 &  0.93 & ~        &   9.2 &        &        &   51 &   468 \\
1: 80 & +75  308 &  7.12 &  0.56 & G0       & +75  309 &  8.24 &  0.75 & G5       &  12.6 &  28.29 &   1.41 &   35 &   446 \\
1: 83 & +22  841 &  8.03 &  0.50 & F5       & +22  842 &  8.33 &  0.46 & F5       &   8.7 &   6.11 &   1.39 &   82 &   719 \\
1: 85 & +03 1049 &  7.02 &  0.51 & F8       & +03 1050 & 10.27 &  0.97 & ~        &  10.0 &  19.63 &   0.70 &   51 &   509 \\
1: 92 & +45  762 &  7.70 &  0.61 & G0V      & +45  761 &  9.17 &  0.92 & K0V      &  29.0 &  26.43 &   1.15 &   38 &  1097 \\
1: 94 & -00 1234 &  9.14 &  0.57 & G0       & -00 1235 & 10.03 &  0.74 & ~        &  39.5 &  13.83 &   1.47 &   72 &  2855 \\
1: 96 & +05 1272 &  7.25 &  0.59 & G1V      & +05 1273 &  8.41 &  0.80 & G5       &  26.3 &  28.10 &   0.92 &   36 &   936 \\
1:104 & +03 1253 &  8.43 &  0.45 & F5       & +03 1252 &  8.31 &  0.54 & F5       &  11.7 &   9.82 &   2.44 &   90 &  1051 \\
1:113 & +10 1224 &  8.85 &  0.92 & G9V      & +10 1223 &  8.76 &  0.86 & G5       &  14.0 &  38.97 &   6.84 &   26 &   360 \\
1:116 & +10 1276 &  8.09 &  1.13 & G5       & +10 1275 & 10.44 &  0.96 & G0       &  18.4 &   7.16 &   1.68 &   75 &  1373 \\
1:117 & +20 1142 &  7.60 &  0.56 & G0       & +19 1017 &  8.40 &  0.70 & G5       &  30.4 &  23.91 &   0.59 &   42 &  1271 \\
1:118 & +06 1268 &  7.64 &  0.44 & F5       & +06 1269 & 10.53 &  0.97 & F5       &  23.4 &  16.66 &   0.67 &   60 &  1404 \\
1:131 & +13 1081 &  9.38 &  0.48 & F8       & +13 1080 &  8.80 &  0.60 & F8       &  17.5 &  10.12 &  11.59 &   95 &  1665 \\
1:138 & +07 1509 &  8.70 &  0.44 & F8       & +07 1510 & 10.15 &  0.75 & G3V      &   7.8 &   9.15 &   1.56 &  109 &   848 \\
1:142 & +53  801 &  6.50 &  0.41 & F6V+...  & +53  800 &  7.95 &  0.58 & F2       &  12.9 &  28.29 &   1.00 &   35 &   456 \\
1:145 & +03 1496 &  6.49 &  0.80 & K0IV     & +03 1497 &  7.53 &  1.02 & K2V      &  28.3 &  56.21 &   0.67 &   18 &   504 \\
1:148 & +14 1209 &  6.20 &  0.57 & G0V      & +14 1208 &  9.22 &  1.14 & G5       &  15.6 &  42.32 &   1.20 &   24 &   370 \\
1:155 & +11 1358 &  8.91 &  0.53 & G5       & +11 1357 & 11.30 &  0.90 & G5       &   8.7 &  10.30 &   1.55 &   97 &   846 \\
1:161 & +19 1174 &  8.22 &  0.71 & G6V      & +19 1175 &  8.43 &  0.76 & G5       &  73.7 &  25.20 &   0.84 &   40 &  2922 \\
1:164 & +55  818 &  7.82 &  0.49 & F8       & +55  819 &  8.35 &  0.57 & F8       &  22.0 &  19.42 &   1.58 &   51 &  1134 \\
1:165 & +53  844 &  8.03 &  0.88 & K0V+K1v. & +53  843 &  8.18 &  0.86 & K0       &  12.7 &  30.61 &   2.13 &   33 &   415 \\
1:167 & +09 1498 &  7.48 &  0.62 & G0       & +09 1497 &  9.64 &  0.98 & K2       &  23.4 &  18.04 &   0.85 &   55 &  1298 \\
1:182 & +24 1359 &  8.16 &  0.68 & G9V      & +24 1360 &  8.59 &  0.76 & K0V      &  39.0 &  27.74 &   0.74 &   36 &  1407 \\
1:184 & -02  789 &  7.68 &  0.51 & F8       & -02  788 & 10.02 &  0.65 & ~        &   8.9 &  18.03 &   0.94 &   55 &   491 \\
1:186 & +02 1666 &  7.06 &  0.77 & G5       & +02 1667 &  7.36 &  0.82 & G5       &  26.7 &  63.35 &   0.94 &   16 &   421 \\
1:192 & +39 1357 &  7.78 &  0.67 & G5V      & +39 1356 &  9.15 &  0.92 & K1III$^a$&  56.2 &  21.74 &   0.80 &   46 &  2585 \\
1:203 & +48 1134 &  7.63 &  0.41 & F5       & +48 1133 &  9.40 &  0.49 & F5       &  19.7 &  11.07 &   1.12 &   90 &  1778 \\
1:213 & +19 1419 &  6.68 &  0.68 & G5V      &l+19 1420 &  7.53 &  0.73 & G7V      &  23.7 &  34.20 &   1.78 &   29 &   694 \\
1:215 & +62  862 &  8.04 &  0.50 & F7V      & +62  861 &  8.74 &  0.65 & B0       &  15.8 &  13.92 &   2.50 &   72 &  1132 \\
1:218 & +10 1798 &  7.13 &  0.49 & F8V      & +10 1799 &  8.01 &  0.58 & G5       &  13.1 &  23.00 &   3.82 &   43 &   571 \\
1:223 & +43 1277 &  7.47 &  0.67 & G2III$^b$& +43 1276 &  9.62 &  1.06 & K        &  40.0 &  31.12 &   0.59 &   32 &  1286 \\
1:224 & +80  347 &  6.58 &  0.67 & G0IV-Vv+ & +80  348 &  7.31 &  0.79 & G5       &  31.3 &  46.03 &   0.29 &   22 &   680 \\
1:225 & +39 1516 &  6.77 &  0.91 & K2V      & +39 1515 &  7.56 &  0.97 & K3V      & 121.8 &  44.76 &   0.42 &   22 &  2721 \\
1:227 & +36 1363 &  7.80 &  0.43 & F5       & +36 1364 &  7.40 &  1.10 & F5       &  14.8 &   7.17 &   5.36 &   63 &   931 \\
1:229 & +36 1372 &  7.57 &  0.48 & F5       & +36 1371 &  8.73 &  0.56 & ~        &  30.0 &  12.73 &   0.55 &   79 &  2357 \\
1:236 & +46 1174 &  7.91 &  0.80 & G5       & +46 1173 & 10.06 &  0.75 & G5       &  93.7 &  11.95 &   0.51 &   84 &  7840 \\
1:247 & +31 1517 &  8.50 &  0.59 & G1V      & +31 1516 &  9.57 &  0.73 & G8V      &  68.0 &  14.37 &   0.69 &   70 &  4733 \\
1:248 & +04 2132 &  8.70 &  0.60 & G5       & +04 2131 & 10.90 &  0.80 & ~        &  24.5 &        &        &   50 &  1227 \\
1:253 & +27 1677 &  3.41 &  0.76 & G5IV     & +27 1676 & 99.99 &-88.74 & ~        &  33.0 & 120.33 &   0.16 &    8 &   275 \\
1:255 & +03 2184 &  7.87 &  0.43 & F5       & +03 2183 & 10.18 &  0.82 & ~        &   9.0 &        &        &   71 &   644 \\
1:257 & +79  491 &  5.68 &  0.46 & K2Vv     & +79  490 &  6.02 &  0.47 & F7       &  19.0 &  15.26 &   3.06 &   17 &   328 \\
1:262 & +06 2406 &  6.92 &  0.43 & F5       & +06 2407 &  9.10 &  0.50 & ~        &   9.8 &  15.17 &   0.82 &   66 &   643 \\
1:266 & +50 1407 &  5.96 &  0.64 & G1.5Vb   & +50 1408 &  6.20 &  0.66 & G3V      &  39.0 &  47.29 &   0.19 &   21 &   824 \\
1:267 & +33 1831 &  4.99 &  0.47 & F7V      & +33 1832 &  8.56 &  1.04 & K6V:     &  25.8 &  47.10 &   0.26 &   21 &   547 \\
1:272 & +20 2208 &  6.49 &  0.36 & F5IV     & +20 2207 &  8.78 &  0.70 & F2       &  11.2 &  22.25 &   0.64 &   45 &   502 \\
1:273 & +57 1348 &  8.67 &  0.43 & F8       & +57 1349 &  9.89 &  0.74 & ~        &   8.0 &        &        &   78 &   624 \\
1:275 & +06 2688 &  7.70 &  0.67 & G4IV     & +06 2687 &  8.02 &  0.64 & G4V      &  42.9 &  16.32 &   0.92 &   61 &  2627 \\
1:296 & +57 1471 &  7.40 &  0.46 & F5V      & +57 1472 &  8.76 &  0.62 & ~        &  12.6 &        &        &   58 &   726 \\
1:298 & +82  650 &  6.98 &  0.45 & F6IV-V   & +82  651 &  7.43 &  0.63 & F5       &  13.4 &  33.47 &   2.12 &   30 &   401 \\
1:306 & +56 1594 &  7.62 &  0.65 & G0       & +56 1595 &  9.67 &  0.79 & ~        &  17.0 &        &        &   40 &   684 \\
1:314 & +47 2022 &  7.14 &  0.61 & G5       & +47 2021 &  7.93 &  0.88 & G5Ve     &  15.1 &  40.59 &   1.88 &   25 &   373 \\
1:321 & +05 3406 &  7.73 &  0.59 & F8       & +05 3405 &  8.50 &  0.73 & F8       &  15.0 &        &        &   61 &   908 \\
1:324 & +24 2585 &  7.87 &  0.61 & G5       & +24 2584 &  9.41 &  0.71 & G0       &   9.4 &  30.00 &   7.00 &   56 &   524 \\
\hline
\end{tabular}
\flushleft \bf
$^a$ the luminosity class is ignored, since it doesn't match the magnitude and the proper motion of the star; see explanations in the text. \\
$^b$ the luminosity class is ignored, since it doesn't match the magnitude and the trigonometric parallax of the star; see explanations in the text.
\end{table*}

%=====================================================================================================================

\section{Statistical properties of the wide binaries}
\label{sec:stat}

%-----------------------------------------------------------------

\subsection{The parent population of the bias-controlled sample of MS wide binaries}
\label{sec:parent}

Our aim is to derive the frequency of MS WBs as a function of the projected separation. We have a sample
of MS WBs with known separations, but we must take into account the selection effects. For that
purpose, we need to select all the stars in the AGK2/3 which could be primary component of a MS wide binary of the
BC sample. This selection is called the parent population of the MS WBs hereafter.

We apply to the stars of the AGK2/3 the same selection criteria as for the primary components of the BC sample, as summarized
hereafter:
\begin{itemize}
\item
Proper motion larger than 50~mas~yr$^{-1}$, since the selection of wide pairs in H86 concerned only stars above this limit.
\item
Photographic magnitude $m_{pg}$ at least as bright as 9.1~mag.
\item
Position at more than 10 degrees from the centers of the four open clusters considered in Section~\ref{sec:BCMSWBCoravel}.
\item
Spectral type at least as late as F5. When existing, the spectral type in Simbad is preferred to that in the AGK2/3.
When the subclass is missing, the $B-V$ color index is used to estimate it.
\item
Luminosity class close to the MS. The luminosity class is taken from Simbad; when the trigonometric parallax of the star is known,
the luminosity class is estimated from the absolute magnitude. Otherwise, the following procedure is applied: the distance of the star    
is derived, assuming it is a giant; the components $(U,V,W)$ of the spatial velocity of the star are derived from the proper motion,
the distance, and the motion of the Sun. When one of these components is 3 times larger than the standard deviation provided by
\cite{Fiala2000}, it is inferred that the star is a dwarf.
\end{itemize}

A parent population of 6372 stars is thus selected for the BC sample. For each star, the spectral type, the mass, the apparent magnitude,
the distance and the proper motion are known.

%------------------------------------------------------------------------

\subsection{Weighting the MS wide binaries of the bias-controlled sample}

The detection probability of any wide binary with a given separation is derived from the parent population with
a simulation: For each ``parent'' star, the mass of a synthetic companion is generated from a 
mass distribution. When the mass of the companion is larger than that of the parent star, the synthetic companion is
ignored, and another one is generated. The magnitude and the $(B-V)$ colour index of the synthetic companion are then
derived from its mass and from the distance of the parent star. They are used to calculate the apparent photographic 
magnitude, thanks to the following equation, which was obtained from the parent population:

\begin{equation}
m_{pg} = m_V + 1.15 \;(B-V)
\label{eq:mpg}
\end{equation}

The companion receives a weight equal to its probability to be recorded in the AGK2/3, due to its brightness.
For a companion brighter than $m_{pg} = 9.15$~mag, the weight is 1; for a faint star, it is the completeness factor read in Fig.~\ref{fig:probaDet}. 
The weight applies to a companion in the allowed range of separations (the apparent and the projected separations are equivalent,
since the distance is known). The closest separation comes from the sum of the radii of the
images of the stars, in equation~\ref{eq:RMpg}. The maximum separation comes from the proper motion, $\mu$, and from the
selection limit of list~2, which was
$T=\rho/\mu < 3500$~yr. Between these two limits, the weight of the companion is added to each bin of the overall
detection probability of the WBs.
This procedure is repeated 10,000 times for each parent star. The overall detection probability as a function of the
projected separation is finally normalized in order to be 1 at its maximum.
%It is worth noticing that ignoring the synthetic companions
%heavier than the parent star results in assuming that, on the main sequence, the intrinsic binary frequency is largest among 
%early-type stars than among late-type stars, as usually found. 

The detection probability of the WBs, as derived above, depends on the mass distribution of the companions. 
In theory, this distribution may be derived from the BC sample, using the method of the ``nested boxes''
\citep{Halbwachs2000, Eggen04}. Unfortunately, the results are not reliable at all, because most
of the WBs have secondary components only about 1 magnitude fainter than the primary component, due
to selection effects. Therefore, the BC sample
leads to the distribution of mass ratios from 1 to around 0.8 only. We prefer then to chose among
theoretical distributions.
To investigate the
importance of this choice, we tried three different distributions: the constant distribution, the Salpeter distribution, 
$f({\cal M}) \propto {\cal M}^{-2.35}$ \citep{Salpeter}, and a distribution obtained by random pairing, assuming the Initial Mass
Function (IMF) found by \cite{kroupa2013}. This IMF is defined as a power law in three parts:

\begin{equation}
\begin{array}{ll}
\psi ({\cal M}) = 0.2410 \; {\cal M}^{-1.3} & , \; {\cal M} \in [0.07, 0.5] \\
\psi ({\cal M}) = 0.1205 \; {\cal M}^{-2.3} & , \; {\cal M} \in [0.5, 1] \\
\psi ({\cal M}) = 0.1205 \; {\cal M}^{-2.7} & , \; {\cal M} \in [1, 10] \\
\end{array}
\label{eq:kroupaIMF}
\end{equation}

\begin{figure}
\includegraphics[clip=,width=\columnwidth]{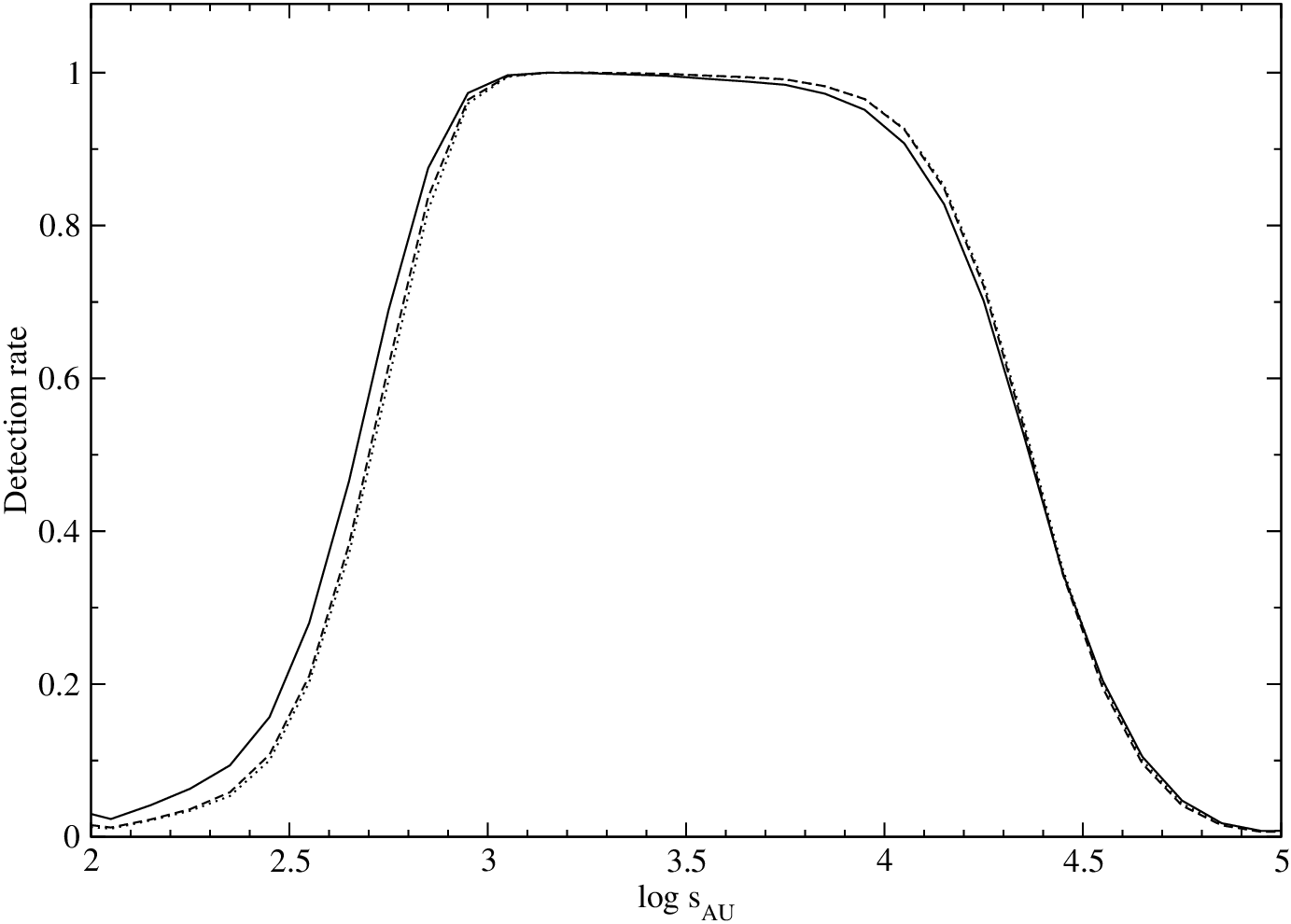}
 \caption{The probability to detect a BC sample binary, as a function of the projected separation.
Three different distribution are assumed for the mass of the companion: dotted line: constant distribution;
dashed line: Salpeter distribution; full line: random pairing from the IMF. To facilitate comparisons,
the maximum probability of each distribution is arbitrarily set to 1.
%; dashes and dots: Random
%pairing from the IMF and constant binary frequency.
}
\label{fig:probDetLogs}
\end{figure}

The resulting probability distributions are presented in Fig.~\ref{fig:probDetLogs}. Surprisingly, the curves obtained from
the three mass distributions are very close. In what follows, the random pairing distribution is assumed.

%--------------------------------------------------------------------------------------------

\subsection{The intrinsic distribution of projected separations}
\label{sec:flogs}

\begin{figure}
\includegraphics[clip=,width=\columnwidth]{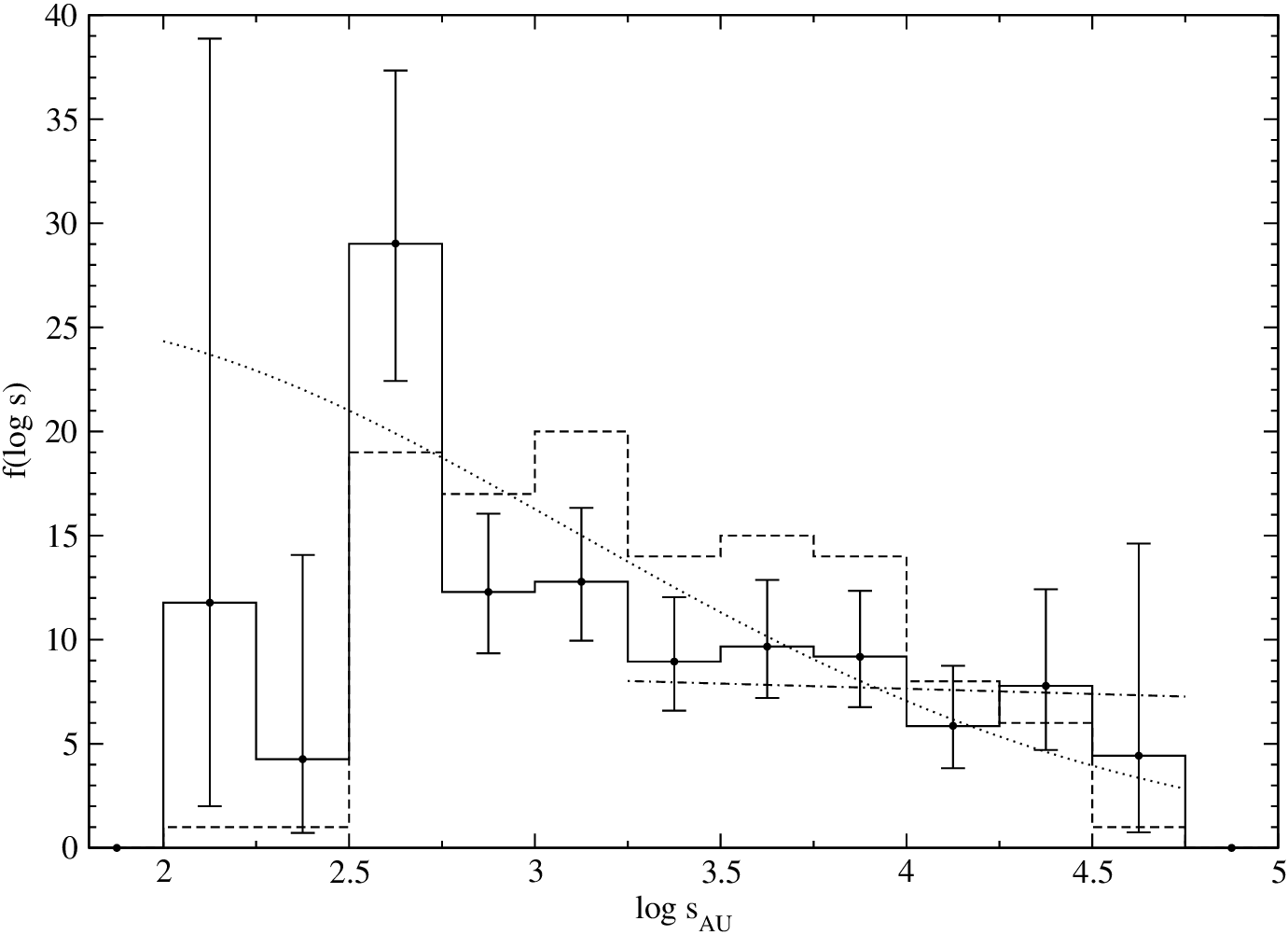}
 \caption{The distribution of the projected separations of the MS wide binaries. The dashed line is the count of the systems
in the bias-controlled sample. The full line is the distribution corrected for selection effects. The dotted line is the
log-normal distribution derived from the parameters of Raghavan (2010), normalized on $\log s$ between 2.5 and 4.5. 
The
dashes and dots correspond to the Power Law distribution $f(a) \propto a^{-1.5}$, for $\log s > 3.25$.
The error bars are derived from the limits of the counts, assuming they obey the Poisson distribution.
}
\label{fig:flogs}
\end{figure}

It is now possible to derive the intrinsic distribution of the projected separations of the MS WBs: For each
of the 116 systems of the BC sample, we obtain a probability from Fig.~\ref{fig:probDetLogs}. We assign
to the system a weight which is the inverse of the probability, and we build the intrinsic distribution of $\log s$
by adding the weights instead of counting the systems. The resulting distribution is in Fig.~\ref{fig:flogs}.
In order to help comparison, it is normalized to obtain the number of WBs in the BC sample.
The error bars are derived from the Poisson distribution: the upper and the lower limits correspond the
the rate parameter providing the actual count or less with a 84~\% probability, or with a 16~\% probability,
respectively. 

%-----------------------
%-----------------------

It appears that the intrinsic distribution of $\log s$ is rather flat for $\log s$ between 2.75 and 4.5, i.e. from about 600 to 30,000~au.
 
We compare our distribution with two distributions frequently found in previous studies, which are the constant distribution, the Power Law
distribution and the
log-normal distribution. The constant distribution of $\log s$ was first found by \cite{Opik1924}, and it was still recently obtained 
by \cite{Longhitano} for WBs, when \cite{LepineBon07} accepted it only until $s=4000$~au.
The Kolmogorov test is used to perform the comparison. The cumulative distribution function of each $\log s$ in our sample is derived,
taking into account the probability detections in Fig.~\ref{fig:probDetLogs}. Since the number of WBs in our sample, $N_{WB}$, is 
sufficiently large, we
consider the product $d_K \sqrt{N_{WB}}$, where $d_K$ is the maximum distance between the cumulative distribution of
$\log s$ for our sample, and that corresponding to \"Opik's law and to the probability detection. We find $d_K \sqrt{N_{WB}} = 1.77$. The probability
to get a so large distance is only 0.3~\%, leading to the conclusion that the constant distribution is inconsistent with our data.

The Power Law distribution $f(a) \propto a^{-\lambda}$
was found by \cite{WassWein} and more recently by \cite{TokoLepine12} for $a$ larger than 2000~au, with $\lambda=1.5$. The latter distribution
is in very good agreement with our
sample for $\log s > 3.25$, since the Kolmogorov test gives $d_K \sqrt{N_{WB}} = 0.7121$, leading to a 66~\% probability.

The log-normal distribution was proposed by \cite{Kuiper35}. Its parameters were accurately derived by \cite{DM91}, and 
slightly ameliorated by \cite{Raghavan10}. From a sample including 87 WBs with periods longer than $1\,000\,000$ days, 
i.e. in the same range of separation as our present sample, they obtained a log-normal distribution
around the mean $\bar{\log P} = 5.03$ when $P$ is expressed in days, and with the standard
deviation $\sigma_{\log P} = 2.28$. Since their study focused on solar-type stars, their distribution of $\log P$ is 
easily converted in a distribution of semi-major axes, $a$. Assuming that the total mass of each system is around 1.5 solar masses,
one obtains another log--normal distribution, around $\bar{\log a} = 1.70$ ($a$ in au), with $\sigma_{\log a} = 1.52$.
Assuming, like them, that $\log s= \log a - 0.13$, the distribution of $\log s$ is log--normal with
the same standard deviation, but around $\bar{\log s} = 1.57$. This distribution is drawn on Fig.~\ref{fig:flogs}.
The Kolmogorov test is used again to check the compatibility of this distribution with our sample. 
We find $d_K \sqrt{N_{WB}} = 0.928$.
The probability to obtain this distance or even more is 34~\%. Therefore, our WBs are in rather good agreement with
the log-normal distribution.

%====================================================================================================================

\section{The multiple systems}
\label{sec:multi}

\subsection{The sample}
\label{sec:SBsample}

\begin{table}
\centering
\begin{minipage}{1\columnwidth}
\caption{\label{tab:multi} The 33 multiple systems among the 80
WBs detected from CORAVEL radial velocities. $P_A$ and $P_B$
are the periods of the components of the WBs when they are SBs.
$P_{\rm WB}$ is an estimation of the period of the WBs. The periods are all
expressed in days. The flag ``f'' is the same as in Table~\ref{tab:WBCora}.
The periods of the SBs considered in Section~\ref{sec:fpSB} are
indicated by an asterisk.}
\begin{tabular}{lcccccc}
\hline
CPM  & $q_A$ & $\log P_A$ & $q_B$ & $\log P_B$ & $\log P_{\rm WB}$ & f\\
\hline
2:  7 &       & 1.399$^*$ &       &           & 7.85 & 3 \\ % SB1O   
2:  8 &       & 1.768$^*$ &       &  ?        & 8.51 & 3 \\ % SB1O SB
1: 18 & 0.996 & 1.173$^*$ &       &           & 7.82 &   \\ % SB2O     
1: 31 &       & 3.225$^*$ &       &           & 8.14 &   \\ %      SB1O
2: 21 &       &           &       & 3.5:      & 8.88 & 2 \\ %      SB1 
2: 33 &       & 1.835$^*$ &       & 3.420$^*$ & 8.72 &   \\ % SB1O SB1O
2: 38 &       &           & 0.746 & 1.752$^*$ & 8.31 &   \\ %      SB2O
2: 40 & 0.620 & 3.167$^*$ &       &           & 7.87 &   \\ % SB2O     
2: 41 &       &           &       & 3.357$^*$ & 8.69 &   \\ %      SB1O
2: 48 &       &           &       & 3.433$^*$ & 8.00 &   \\ %      SB1O
2: 49$^a$ &       & 3.433 &       &           & 8.10 & 4 \\ % SB1O     
1: 93 & 0.857 & 3.176$^*$ &       &           & 8.25 &   \\ % SB2O     
1:112 &       & 3.449$^*$ &       &           & 7.09 &   \\ % SB1O     
1:114 &       & 2.584$^*$ &       &           & 8.11 &   \\ % SB1O     
2: 58 &       & 1.792$^*$ & 0.624 & 3.362$^*$ & 9.38 &   \\ % SB1O SB2O
1:130 &  ?    &  ?        &       & 0.750$^*$ & 8.47 &   \\ % SB2  SB1O
1:141 &       & 1.938     &       &           & 8.72 & 2 \\ % SB1O     
2: 65 &       &           &       & 2.978$^*$ & 8.55 &   \\ %      SB1O
1:156 & 0.913 & 1.855     &       &$>3.6$     & 7.92 & 1 \\ % SB2O SB1 
2: 70 &       & 3.612     &       &           & 9.10 &   \\ % SB1O     
2: 73 &       &           &       & 3.462     & 8.75 & 1 \\ %      SB1O
2: 74 &       &           & 0.957 & 1.289     & 8.82 & 1 \\ %      SB2O
1:175 &       & 3.464     &       &           & 7.60 & 1 \\ % SB1O     
2: 89 &       &$>3.9$     &       &           & 8.68 & 2 \\ % SB1      
2: 92 &       & 0.742$^*$ &       &           & 8.71 & 3 \\ % SB1O   
1:246 &       &           & 0.643 & 0.773$^*$ & 7.83 & 2 \\ %      SB2O
2: 94 &       & 2.974$^*$ &       &           & 8.92 &   \\ % SB1O     
1:258 &       & 3.546$^*$ &       &           & 7.98 &   \\ % SB1O     
1:280 & 0.845 & 2.206$^*$ &       &           & 7.64 &   \\ % SB2O     
1:300 &       &           & 0.933 & 0.968$^*$ & 8.32 &   \\ %      SB2O
1:307 & 0.566 & 0.660$^{b\;*}$ &       &           & 7.82 &   \\ % SB2O    
\hline
\end{tabular}
\flushleft \bf
$^a$ The WB component 2:49A is also 2:48B. \\
$^b$ 1:307A is a triple system; we keep only the shortest period. The long period is 50 yr.
\end{minipage}
\end{table}

Thirty-three WBs have at least one SB component, and the systems 
in Table~\ref{tab:WBCora} includes a total of 37 SBs which were studied in Paper I.
The mass ratios of the double-lined SBs (SB2), $q$, and the logarithms of the periods of the SB orbits are
listed in Table~\ref{tab:multi}.

Despite its moderate size, this sample may be used to investigate some properties of
multiple systems. So, we will see the 
correlation between the periods of WBs and that of SBs, the
ratio between the frequencies of quadruple and triple systems, and 
the period distribution of the SBs.

%-------------------------------------------------------------------

\subsection{Frequency of quadruple systems with respect to the triple systems}

We want to see if the formation of a close binary is due to peculiar conditions that
are common to both components. If this is true, quadruple systems are expected to
be more frequent than when we assume that the probability to have a
SB is the same for any WB components.

To investigate this question, we consider the WBs for which a SB could be
equally detected for both components. We discard the systems belonging to
open clusters ($f=1$), since their selection as WBs is questionable.
The systems containing a late-type
giant or a star earlier than F5 ($f=2$) are discarded, because the former correspond to
long-period SB, and the latter are difficult to measure with CORAVEL. 
The detection of SBs with unknown periods, or with periods longer than 10 years,
is facilitated when the other WB component is also a SB, since both stars
were then observed more frequently than the others; for that reason, only
WB with SB with $P < 10$~yr, or $\log P < 3.563$ are considered. 
The system 2:49 ($f=4$) is discarded too, since its SB component belongs also
to 2:48.
After this selection, we count 22 SBs among the 53 systems of Table~\ref{tab:WBCora}
which have not $f=1$, 2, or 4, and which are not rejected because the period of the SB component
is longer than 10~yr.
The overall SB proportion is then $p_{SB}=22/(2\times53)$. The expected number
of triple systems is $n_3=53 \times p_{SB} \times (1- p_{SB}) \times 2 = 17.4$,
and the expected number of quadruple systems is $n_4 = 53 \times p_{SB}^2 = 2.3$.
These numbers match very well to the actual counts, which are 18 and 2, respectively.
The fact that the number of quadruple systems is exactly that expected from
independent probabilities firmly supports the hypothesis
that the probability to be a close binary 
with a period shorter than 10~yr is the same for any wide binary
component, regardless of the possible binarity of the other component.
In fact, the excess of quadruple systems found by
\citet{Toko14} and \citet{Riddle15} is due to subsystems with periods
longer than 10~yr.

%-------------------------------------------------------------------

\subsection{Properties of the tertiary components}

We consider now the 18 triple systems of the previous section, in order to
see if the single component of the WB is less massive than the SB, as
suggested in the past by \cite{Toko08}. We count 12 SBs as the ``A'' component of the WB,
and 6 SBs as the ``B'' component. However, the ``A'' component is sometimes the
brightest one because it is a close binary: it appears in Table~\ref{tab:WBCora} that
the spectral types of the ``B'' components of the WBs 1:18, 1:114, 2:92 and 1:280  are
earlier than those of the ``A'' components. Therefore, when we define the
primary component of a WB as the heavier star of the triple system, we count
8 SBs among the primary components, but 10 among the secondaries. Therefore, 
we agree with the recent result
of \cite{Toko15} that the frequency of SBs is the
same for any component of the WBs.

\cite{Toko08} also pointed out that, in a triple system, the close binary is
usually the heaviest component of the WB. This applies also to our sample.
However, this may be due to a selection effect: as noticed by
\cite{LucyRicco79}, a binary star is much fainter than a single star with the
same total mass, and our sample is delimited by apparent magnitude. 
Therefore, we can't draw any conclusion about that question.

%-------------------------------------------------------------------

\subsection{Correlation between the SB periods and the WB periods}

\begin{figure}
\includegraphics[clip=,width=\columnwidth]{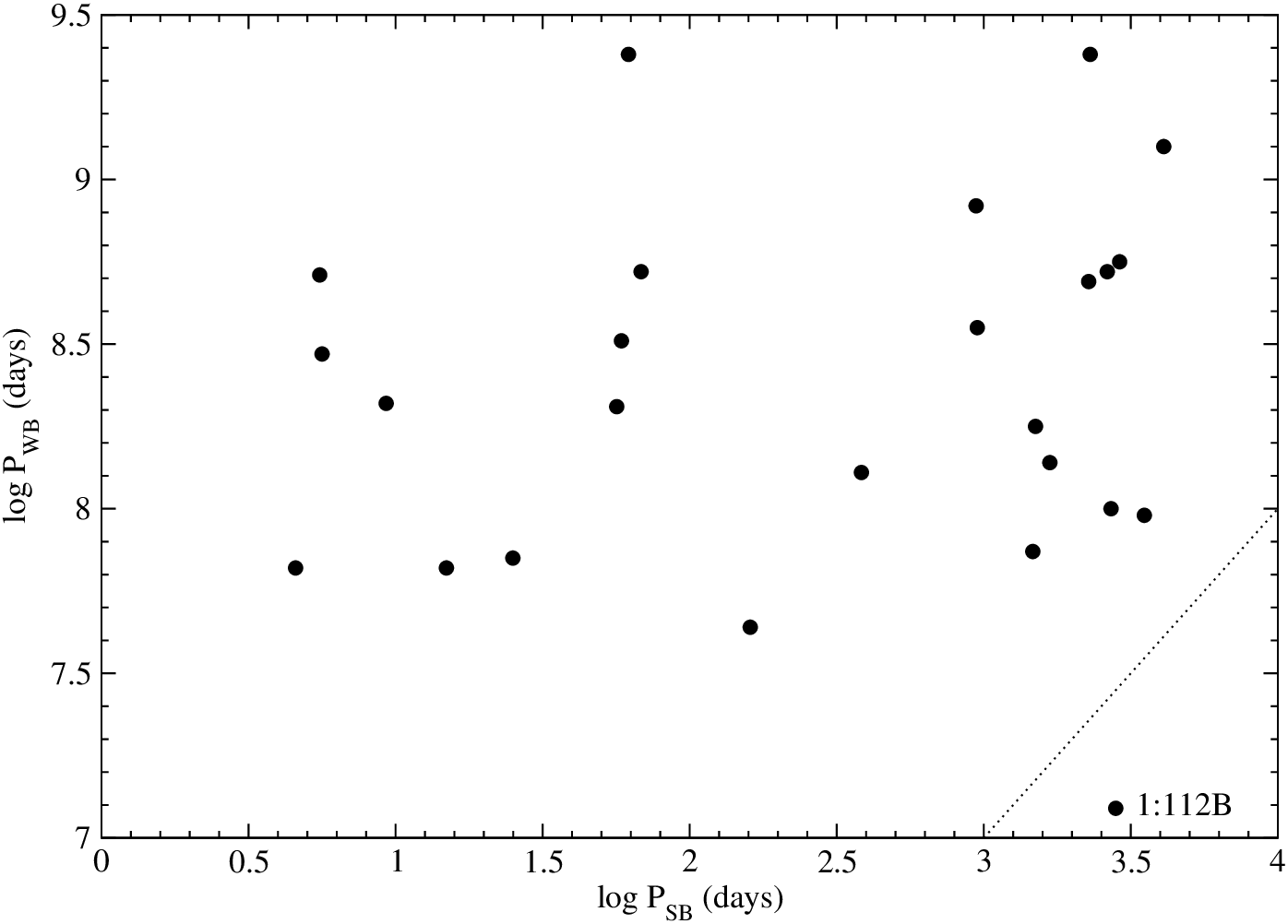}
 \caption{The period-period diagram of the 25 components of MS field WBs 
as late as F5 which are SBs. The dotted line is the limit $\log P_{WB} >
\log P_{SB} + 4$.
}
\label{fig:logPSBlogPWB}
\end{figure}

The periods of the WBs are derived from our calculations in Section~\ref{sec:flogs}. 
They are included in Table~\ref{tab:multi}. For homogeneity, we consider a possible
relation between the periods of the SBs and that of the WB, for the MS stars at least
as late as F5 which are not in open clusters. Discarding the SBs of Table~\ref{tab:multi}
with $f= 1$ or 2, one obtains the $\log P_{SB}$ vs $\log P_{WB}$ diagram
in Fig.~\ref{fig:logPSBlogPWB}. No obvious correlation appears on this diagram.
The SBs with the longest periods are components of WBs with relatively short as well as
with very long periods.

In fact, this result is not in contradiction with \cite{Toko08}, who found that $\log P_{WB}$ is
usually less than $\log P_{SB} + 4$. In Fig.~\ref{fig:logPSBlogPWB}, only
one WB has a period shorter than this limit: 1:112B, in the low right corner of the figure,
should have $\log P_{WB}$ larger than 7.4 instead of equal to 7.09. The limitation of
the SBs to periods shorter than 10 years makes our sample unsuitable to check
Tokovinin's hypothesis.

%----------------------------------------------------------------------------------

\subsection{Distribution of periods}
\label{sec:fpSB}

Finally, we want to compare the distribution of periods of SBs belonging to WBs to
that of close binaries as a whole. For that purpose, we consider the log-normal
distribution found by \cite{Raghavan10}, that we have already used in Section~\ref{sec:flogs}. 
We select the SBs 
with a MS primary component at least as late as F5, whatever the type of the wide
companion is, except those belonging to an open cluster. 
These SBs are indicated with an asterisk in Table~\ref{tab:multi}. A comparison to
the log-normal distribution until $P<10$~yr is made with the two-side version of the test of Kolmogorov. The major
distance between the cumulative distributions of periods is 0.168. For a sample of 24
objects, the probability to get a maximum distance even larger is 48~\% when both distributions are equal
in reality. Therefore, the log-normal distribution is accepted for the SBs belonging to
WBs.

\section{Conclusion}
\label{sec:conclusion}

We have carried out an observation programme, essentially with CORAVEL, in order to select a sample of WBs. 
We obtained a total amount of 3748 RV measurements. These observations were
used to derived the orbital elements of 52 SBs (in Paper I), and to select 80 WBs with compatible RVs,
parallaxes, and proper motion. Adding some close WBs from list~1 of H86, we finally obtained a sample of 116 
main-sequence WBs with spectral types
between F5 and M0. At the same time, 39 optical pairs were found in list~2 of H86.

The sample of main-sequence WBs was used to derived the distribution of the separations. The constant distribution, or ``\"Opik's law''
was rejected, but the log-normal distribution corresponding to the period distribution of \cite{Raghavan10} was accepted. The
Power Law of \cite{TokoLepine12}, $f(a) \propto a^{-1.5}$, was widely accepted for $\log s > 3.25$. 
\citet{Weinberg90} derived from theoretical calculations
that the distribution of $a$ is not necessarely ending with a break. No evidence for a break is visible in 
\cite{TokoLepine12} until $s=64,000$~au, neither in our data.
However, this is due to the selection of the sample: the end of the distribution could be found only by extending the search for WBs
beyond the limit of $T$ which was applied in order to have a proportion of genuine WBs as large as 60~\%. This will be possible
when the Gaia satellite \citep{gaia} will provide very accurate parallaxes, proper motions and radial velocities, and when the apparent separation
will no longer be a basic criterion in the selection of WBs.

The SBs found in Paper I were used to investigate the statistical properties of the multiple systems.
No correlation was found between the SBs and the WB: The proportion of quadruple systems was compatible with a random distribution
of SBs, and the SB components were equally found around the heaviest as around
the lightest WB components. The first of these two results is in contrast to that of \citet{Toko14} and \citet{Riddle15}, but their statistics include
a majority of inner binaries with periods larger than 10 years. Therefore, it seems that the environment of WBs may
facilitate the formation of binaries with periods longer than a certain limit. Close binaries with periods shorter than
10 years are not concerned by this effect.

No correlation was found between the periods of the SBs and those of the WBs, but this is due to the
limitation of our sample to SBs with periods until 10 years, which is much less than the periods of the WBs.

Finally, the distribution of periods of the SBs is also compatible with that of \cite{Raghavan10} until $P=10$~yr.

%__________________________________________________________________________________________

\section*{Acknowledgements}

It is a pleasure to thank Dr Thomas Maschberger for relevant informations about the IMF and the mass ratio
distribution. We had a fruitful e-mail exchange with Prof. Leon Lucy about the selection effects
affecting the binary stars.
We have benefitted during the entire period of these observations from the support of the Swiss National
Foundation and Geneva University. We are particularly grateful to our technicians Bernard Tartarat, Emile
Ischi and Charles Maire for their dedication to that experiment for more than 20 years.
We made use of Simbad, the database of the Centre de Donn\'ees astronomiques de
Strasbourg (CDS).%

%===========================================================================================
%\online

\appendix

\section[]{Detection limit of close companions in the AGK2/3}
\label{sec:DetLim}

\begin{table*} 
\caption{\label{tab:rho-mpg} Search of the detection limit as a function of the photographic 
magnitude of the primary component, $m_{pg}$, and of the apparent separation, $\rho$.
 .}
%\scriptsize
\begin{tabular}{c|rc|rc|rc|rc|rc|rc}
\hline
         & \multicolumn{12}{c}{$\rho$~(arcsec)} \\
         & \multicolumn{2}{c}{$<$~5} & \multicolumn{2}{c}{5 -- 10} & \multicolumn{2}{c}{10 -- 20} & \multicolumn{2}{c}{20 -- 40} & \multicolumn{2}{c}{40 -- 80} & \multicolumn{2}{c}{$>$~80} \\
$m_{pg}$ &  $n_i$ & $f_i$ &  $n_i$ & $f_i$ &  $n_i$ & $f_i$ &  $n_i$ & $f_i$ &  $n_i$ & $f_i$ &  $n_i$ & $f_i$ \\
\hline
$>$~10  & 3 & 1.00 & 29 & 1.00 & 33 & 1.00 & 16 & 1.00 & 13 & 1.00 & 66 & 1.00 \\
9 -- 10 & 2 & 0.40 & 17 & 0.37 & 25 & 0.43 & 19 & 0.54 & 10 & 0.43 & 107& 0.62 \\
8 -- 9  & 0 & 0    & 17 & 0.27 & 19 & 0.25 & 15 & 0.30 & 22 & 0.49 & 137& 0.44 \\
7 -- 8  & 1 & 0.17 &  4 & 0.06 & 13 & 0.14 & 15 & 0.23 & 19 & 0.30 & 83 & 0.21 \\
6 -- 7  & 0 & 0    &  0 & 0    &  7 & 0.07 &  4 & 0.06 &  4 & 0.06 & 26 & 0.06 \\
5 -- 6  & 0 & 0    &  0 & 0    &  4 & 0.04 &  6 & 0.08 &  2 & 0.03 & 27 & 0.06 \\
4 -- 5  & 0 & 0    &  0 & 0    &  0 &  0   &  1 & 0.01 &  1 & 0.01 & 9  & 0.02 \\
\hline
\end{tabular}
\end{table*}

We need the minimum separation still allowing complete detection of the
components. We consider the 780 pairs with $T < 10,000$~years selected in the
course of the preparation of H86. The distribution of these pairs
according to the apparent separation and the photographic magnitude of
the primary component is presented in Table~\ref{tab:rho-mpg}. Each count, $n_i$ is
followed with its rate, $f_i$, defined as the ratio:

\begin{equation}
f_i = \frac{n_i}{ \sum_{j \leq i} n_j}
\end{equation} 

\noindent
where $j$ is the index of the rows from the top of the table until $i$ included.
For a given separation, it is easier to detect a companion when the primary is faint 
than when it is bright. Therefore, the table is presented with rows in increasing order 
from the faintest until the brightest magnitudes. We read each row from right to left: as long as
the separation is larger than the detection limit, the rate $f_i$ is expected to be constant 
(the variations are only due to Poisson noise); when the detection is not complete, $f_i$
is decreasing. We infer from Table~\ref{tab:rho-mpg} that the radius of the image of each star,
in arcseconds,
is a bit less than that given by the following formula:

\begin{equation}
\label{eq:RMpg}
\log R_{\rm arcsec} = \frac{13 - \min(m_{pg},10)}{7} 
\end{equation}

We assume that the detection is complete when the separation $\rho$ is larger than the sum of the
radii of the components.

%\section[]{The photographic magnitudes in the AGK2/3}
%\label{sec:photmag}

\label{lastpage}

\end{document}